\documentclass[aps,prl,reprint,superscriptaddress,floatfix,amssymb]{revtex4-2}
\usepackage{xcolor,amsmath,url,graphicx,hyperref}
\usepackage{subcaption}
\usepackage{siunitx}
\usepackage[capitalize]{cleveref}

\begin{document}
\title{Strong Reduction of Quasiparticle Fluctuations in a Superconductor due to Decoupling of the Quasiparticle Number and Lifetime}

\author{Steven A. H. de Rooij}
\email{s.a.h.de.rooij@sron.nl}
\affiliation{SRON - Netherlands Institute for Space Research, Niels Bohrweg 4, 2333 CA Leiden, The
	Netherlands}
\affiliation{Faculty of Electrical Engineering, Mathematics and Computer Science, Delft University of
	Technology, Mekelweg 4, 2628 CD Delft, the Netherlands}

\author{Jochem J. A. Baselmans}
\affiliation{SRON - Netherlands Institute for Space Research, Niels Bohrweg 4, 2333 CA Leiden, The
	Netherlands}
\affiliation{Faculty of Electrical Engineering, Mathematics and Computer Science, Delft University of
	Technology, Mekelweg 4, 2628 CD Delft, the Netherlands}

\author{Vignesh Murugesan}
\affiliation{SRON - Netherlands Institute for Space Research, Niels Bohrweg 4, 2333 CA Leiden, The
	Netherlands}

\author{David J. Thoen}
\affiliation{Faculty of Electrical Engineering, Mathematics and Computer Science, Delft University of
	Technology, Mekelweg 4, 2628 CD Delft, the Netherlands}
\affiliation{Kavli Institute of NanoScience, Faculty of Applied Sciences, Delft University of Technology, Lorentzweg 1, 2628 CJ Delft, The Netherlands}

\author{Pieter J. de Visser}
\affiliation{SRON - Netherlands Institute for Space Research, Niels Bohrweg 4, 2333 CA Leiden, The
	Netherlands}

	\begin{abstract} 		
		We measure temperature dependent quasiparticle fluctuations in a small Al volume, embedded in a NbTiN superconducting microwave resonator. The resonator design allows for read-out close to equilibrium. By placing the Al film on a membrane, we enhance the fluctuation level and separate quasiparticle from phonon effects. When lowering the temperature, the recombination time saturates and the fluctuation level reduces a factor $\sim$100. From this we deduce that the number of free quasiparticles is still thermal. Therefore, the theoretical, inverse relation between quasiparticle number and recombination time is invalid in this experiment. This is consistent with quasiparticle trapping, where on-trap recombination limits the observed quasiparticle lifetime.
	\end{abstract}
\maketitle

\begin{figure}
	\includegraphics[width=\linewidth]{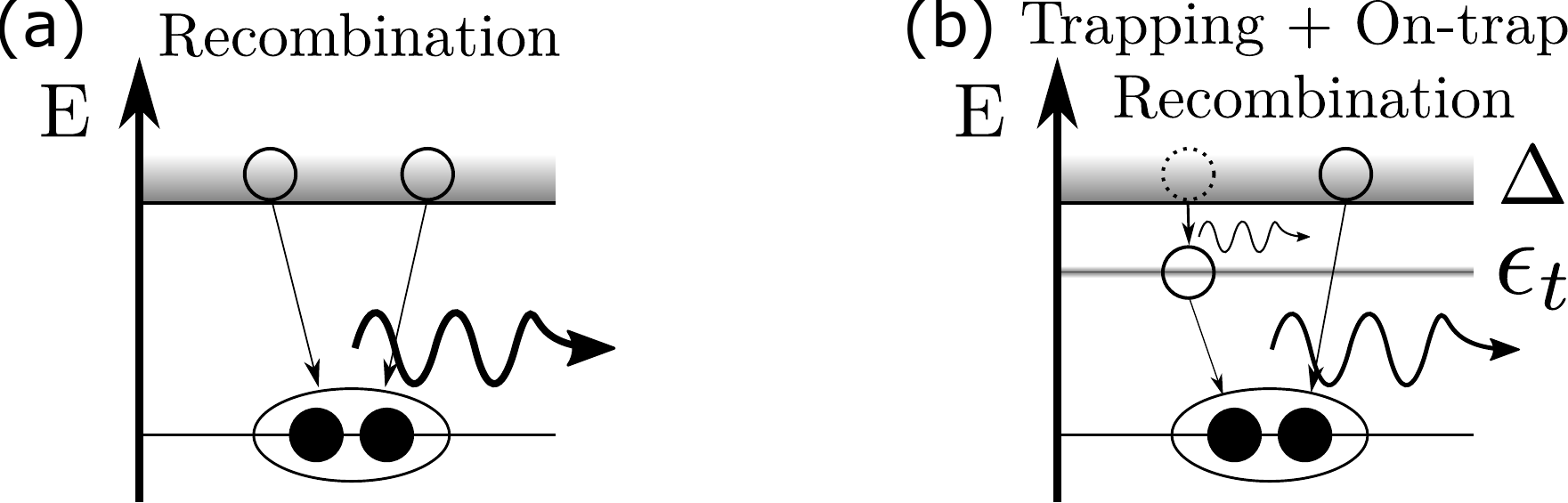}
	\caption{Schematic representations of simple recombination \textbf{(a)} and trapping with subsequent on-trap recombination \textbf{(b)} of two quasiparticles. $\Delta$ is the superconducting energy gap and $\epsilon_t$ the trap energy. Curved arrows represent emitted phonons, with thickness indicating their relative energy.}\label{fig:diagram}
\end{figure}

In a superconductor well below its critical temperature, most electrons are bound together in Cooper pairs, while the number of unpaired electrons, quasiparticles, exponentially decreases with decreasing temperature. The recombination time of quasiparticles is inversely proportional to the quasiparticle number and therefore increases exponentially towards lower temperature. At a constant, finite temperature, the quasiparticle number fluctuates around its average value due to random generation and recombination events \cite{Wilson2004}. In Ref. \cite{deVisser2011}, a measurement of these fluctuations showed that the quasiparticle number and recombination time are indeed inversely proportional, and as a consequence the quasiparticle fluctuation level is \textit{constant} as a function of temperature. Both observations are consistent with direct recombination of free quasiparticles with emission of a phonon \cite{Kaplan1976}, see \cref{fig:diagram}(a). Also in the presence of excess quasiparticles at low temperature, which often occur in  superconducting devices in varying conditions \cite{Martinis2009,Rainis2012,Barends2011,Karatsu2019,Cardani2021,deVisser2014a}, the relation between the number of quasiparticles and their recombination time is maintained \cite{deVisser2011}.  \\ 

In this Letter we show experimentally that the intimate relation between the free quasiparticle number and recombination time is broken when quasiparticles are first trapped and then recombine as depicted in \cref{fig:diagram}(b).\\
Impurities and disorder in superconductors are known to reduce the recombination time at low temperatures \cite{Barends2009a,Gao2012} and can fundamentally change the relation between quasiparticle number and recombination time. The low-temperature recombination time is typically limited to tens of microseconds \cite{Gao2012,Barends2008} to milliseconds \cite{deVisser2011,Fyhrie2020}, depending on the material. However, these experiments only measure the recombination time from a non-equilibrium pulse decay, or suffer from excess quasiparticles. Moreover, in tunnel junction devices, including qubits, only non-equilibrium, local quasiparticle properties can be measured directly. Understanding the reduced recombination time at low temperatures and close to thermal equilibrium, and in particular its relation to the quasiparticle density, has been hindered by the lack of a sensitive probe of quasiparticle dynamics.\\

Here, we measure quasiparticle fluctuations in a small Al volume (\SI{27}{\cubic\micro\meter}), embedded in a NbTiN superconducting resonator. This design allows for read-out at low microwave powers, minimizing the creation of excess quasiparticles \cite{deVisser2012}. By placing the Al volume on a membrane, we enhance the fluctuation level and separate quasiparticle from phonon effects.\\
We observe a strong reduction in the quasiparticle fluctuation level by a factor $\sim$100 when lowering the bath temperature from \SIrange{350}{200}{\milli\kelvin}, together with a saturation of the recombination time. Hence, the inverse proportionality between quasiparticle number and recombination time must be broken and the recombination time is not limited by excess quasiparticles as we observed before \cite{deVisser2012,deVisser2011}. Together, our observations are in qualitative agreement with quasiparticle trapping and subsequent on-trap recombination.\\
Our methodology is well suited to better understand the quasiparticle dynamics in devices where quasiparticle trap structures \cite{Riwar2016,Court2008} and vortices \cite{Wang2014} are introduced deliberately, to reduce the excess quasiparticle density in critical regions of a device.\\

In thermal equilibrium and at low temperatures ($T\ll\Delta/k_B$), the number of quasiparticles in a superconducting volume $V$, is given by \cite{Tinkham2004},
\begin{equation}\label{eq:Nqp}
	N_{qp}(T) = 2VN_0\sqrt{2\pi k_BT\Delta}e^{-\Delta/k_BT},
\end{equation}
where $N_0$ is the single spin density of states at the Fermi-level (we use $N_0= \SI{1.72e4}{\micro\electronvolt^{-1}\micro\meter^{-3}}$, for Al), $k_B$ is the Boltzmann constant and $\Delta=1.76k_BT_c$ is the superconducting gap energy, with $T_c$ the critical temperature. The intrinsic quasiparticle lifetime with respect to recombination (hereafter called 'the quasiparticle lifetime') is given by \cite{Kaplan1976},
\begin{equation}\label{eq:tqp}
	\tau_{qp}(T) = \frac{\tau_0 N_0 V (k_BT_c)^3}{2\Delta^2 N_{qp}(T)} = \frac{V}{RN_{qp}(T)},
\end{equation}
where $\tau_0$ is a material dependent characteristic time for the electron-phonon coupling. For Al, we take $\tau_0 = \SI{0.44}{\micro\second}$ \cite{Kaplan1976}. In the last equality, all the material dependent parameters are combined into the recombination constant, $R$.\\
In experiments, the relaxation of an ensemble of quasiparticles is typically probed. As the recombination of two quasiparticles into a Cooper-pair is a pair-wise process and the emitted phonon can subsequently break a Cooper-pair \cite{Rothwarf1967}, the \textit{apparent} quasiparticle lifetime is given by,
\begin{equation}\label{eq:tqpstar}
	\tau_{qp}^* = \tau_{qp}(1+\tau_{esc}/\tau_{pb})/2.
\end{equation}
Here, $\tau_{esc}$ is the phonon escape time, $\tau_{pb}$ is the phonon pair-breaking time and the factor in parentheses is called \textit{the phonon trapping factor}. \Cref{eq:tqpstar} is valid when, $\tau_{esc},\tau_{pb}\ll\tau_{qp}$, which is typically the case \cite{Kaplan1979}. We take $\tau_{pb}=\SI{0.28}{\nano\second}$ for Al \cite{Kaplan1976}. $\tau_{esc}$ can experimentally be tuned \cite{Rostem2018,Puurtinen2020}, for instance with the use of a membrane, which we use here to distinguish phonon effects from intrinsic quasiparticle processes.\\
Fluctuations in the quasiparticle number occur randomly. Starting from a master equation approach \cite{Wilson2004}, the Power Spectral Density (PSD) of these fluctuations can be calculated to be,
\begin{equation}\label{eq:Sn}
	S_{N_{qp}}(\omega) = \frac{4\tau_{qp}^*N_{qp}}{1+(\omega\tau_{qp}^*)^2}.
\end{equation}
This is a Lorentzian spectrum, with a constant level with temperature (since $\tau_{qp}^*\propto 1/N_{qp}$) and a roll-off frequency of $\omega=1/\tau_{qp}^*$.\\
\begin{figure*}
	\includegraphics[width=\linewidth]{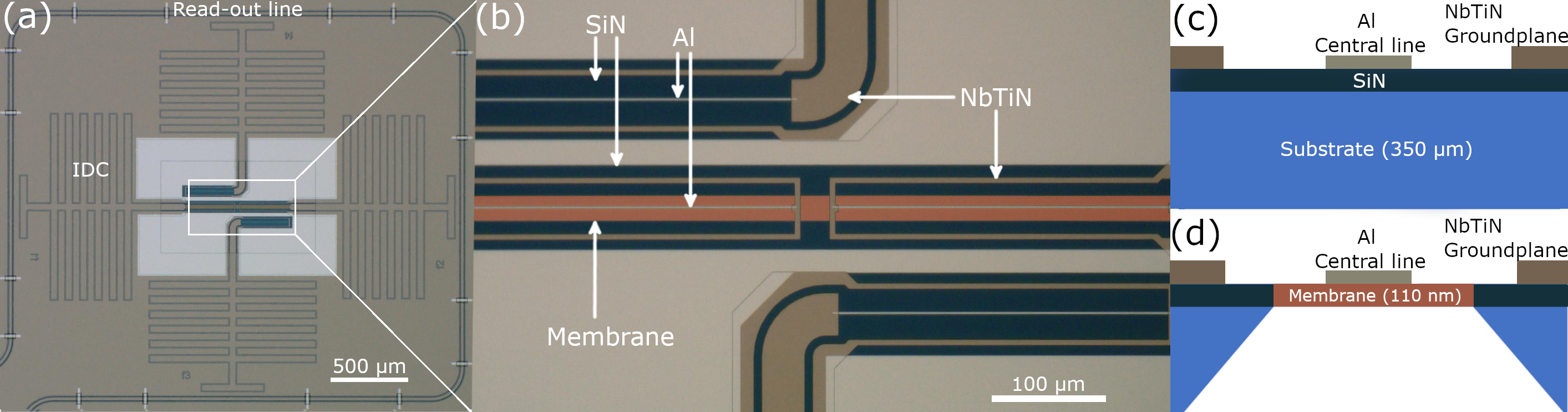}
	\caption{\textbf{(a)}: Micrograph of the resonators studied. \textbf{(b)}: Zoom-in on the Al inductive parts of the resonators, of which two are on a \SI{110}{\nano\meter} thick SiN membrane (highlighted in orange). \textbf{(c,d)}: Schematic cross sections (not to scale) of the inductive (Al) part of a substrate and membrane resonator, respectively. The CPW dimensions are such, that only the Al central line is suspended by the membrane.}\label{fig:LT165}
\end{figure*}

The device under study is shown in \cref{fig:LT165}. Four NbTiN-Al hybrid \cite{Yates2011} resonators are capacitively coupled to a read-out line patterned around them. The capacitive part is a NbTiN double sided interdigitated capacitor (IDC) \footnote[2]{J. J .A. Baselmans et al., in preparation.}\cite{Noroozian2009a} and the inductive part is a NbTiN co-planar waveguide with an Al central line. For details see the Supplemental Material \footnotemark[1].\\
The resonator response is only sensitive to quasiparticle - Cooper-pair fluctuations within the Al ($V=\SI{27}{\cubic\micro\meter}$ and $T_c=\SI{1.26}{\kelvin}$), as we measure at $T\ll T_{c,NbTiN} = \SI{15.6}{\kelvin}$ and the current density is much higher in the Al section due to the IDC design. Furthermore, as $\Delta_{Al}\ll\Delta_{NbTiN}$, quasiparticles are confined to the Al volume.\\
The resonators are mounted in a pulse tube pre-cooled adiabatic demagnetization refrigerator, surrounded by a cryophy and a superconducting magnetic shield, and isolated from stray light by a 'box-in-a-box' configuration \cite{Baselmans2012a}. The forward microwave transmission, $S_{21}$, is recorded at an on-chip read power, $P_{read}=\SI{-99}{\decibel m}$. In the following, we present the results for two resonators, of which one has its sensitive volume on a membrane and one on the substrate, see \cref{fig:LT165}(c,d).\\
We translate the complex $S_{21}$ to an amplitude ($A$) and phase ($\theta$) to distinguish changes in dissipation ($A$) and kinetic inductance ($\theta$). These variables are defined relative to the resonance circle, which is measured for each temperature before the fluctuation measurement. The responsivity is given by $dX/dN_{qp} = 2\alpha_kQ\kappa_X/V$, where $X$ is either $A$ or $\theta$ and $\alpha_k$ is the fraction of kinetic inductance over the total inductance. $\kappa_X$ describes the change in complex conductivity with respect to a change in quasiparticle density, which is only weakly dependent on temperature \cite{Gao2008}. $Q$ is the loaded quality factor and measured to be \SI{4.3e4}{} and \SI{3.5e4}{} at \SI{50}{\milli\kelvin}, for the membrane and substrate resonator, respectively. The resonators are designed to have the same volume and kinetic inductance fraction.\\ 
We measure \SI{40}{\second} time streams at \SI{50}{\kilo\hertz} of $A$ and $\theta$, at temperatures ranging from \SIrange{50}{400}{\milli\kelvin}. We filter pulses caused by cosmic rays \cite{Karatsu2019} or other external sources \footnote[1]{See Supplemental Material for all power spectral densities, device fabrication details, pulse rejection, responsivity calculations and influence of Al geometry and read power.} and calculate the cross-PSD, $S_{A,\theta}(\omega)$. By using the cross-PSD, we extract the dissipation - kinetic inductance (i.e., quasiparticle - Cooper-pair) fluctuations and suppress uncorrelated noise sources such as amplifier and TLS noise \cite{deVisser2012}. We determine the spectrum of the quasiparticle fluctuations via,
\begin{equation}\label{eq:Sn_exp}
	S_{N_{qp}}(\omega) = S_{A,\theta}(1+(\omega\tau_{res})^2)\left(\frac{dAd\theta}{dN_{qp}^2}\right)^{-1},
\end{equation}
where $\tau_{res}=Q/\pi f_0$ is the resonator ring time ($f_0$ is the resonance frequency), typically a few \si{\micro\second}. This implies $(\omega\tau_{res})^2\ll1$ for frequencies below \SI{100}{\kilo\hertz} and we therefore neglect this factor. The last factor in \cref{eq:Sn_exp} is the multiplication of the amplitude and phase responsivities, which we determine from a measurement of $S_{21}(\omega;T)$, for $T>\SI{250}{\milli\kelvin}$ \footnotemark[1].\\

\begin{figure*}[t]
	\includegraphics[width=.9\linewidth]{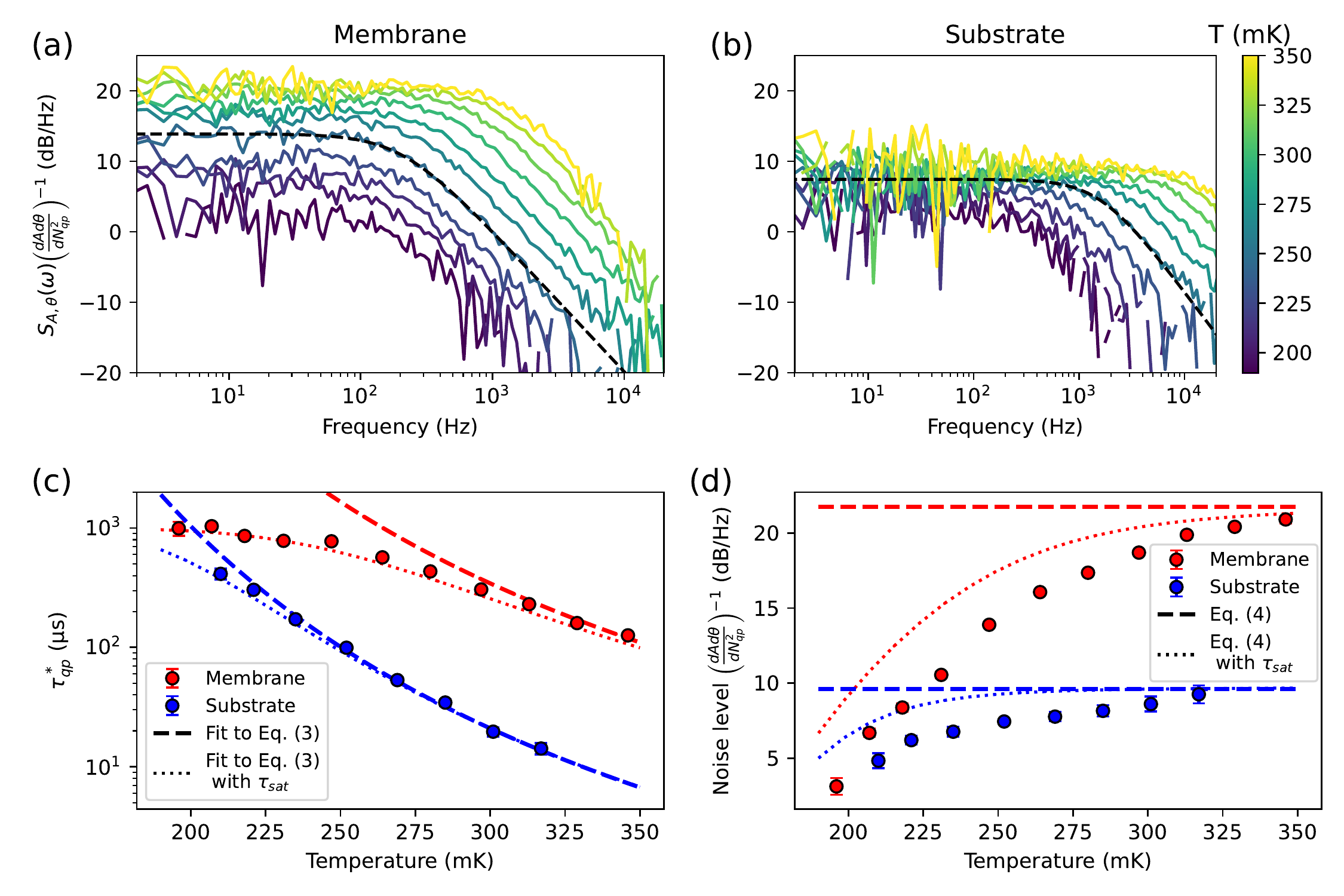}

	\caption{\textbf{(a,b)}: Measured quasiparticle fluctuations for a resonator on membrane (a) and on substrate (b). The quasiparticle fluctuations are determined from the measured cross-PSD, using the measured responsivities and \cref{eq:Sn_exp}. The measurements are preformed at \SI{-99}{\decibel m} read-out power. The dashed black lines give examples of the Lorentzian fits (\cref{eq:Sn}).\textbf{(c,d)}: Lifetimes (c) and noise levels (d) from Lorentzian fits (\cref{eq:Sn}) to the spectra in (a,b). The errorbars indicate statistical uncertainties from the fitting procedure. Only fits with a relative fitting uncertainty lower than \SI{13.5}{\percent} in lifetime are displayed. The dashed lines in the lifetime plot are fits to \cref{eq:tqp,eq:tqpstar} for temperatures \SI{\geq 300}{\milli\kelvin}, with $\tau_{esc}$ as free parameter. The dashed lines in the level plot are calculated from \cref{eq:Sn}, with \cref{eq:Nqp,eq:tqpstar} and the same $\tau_{esc}$. The dotted lines are calculated in the same way, but with $\tau_{qp}^*\rightarrow\left((\tau_{qp}^*)^{-1}+(\tau_{sat})^{-1}\right)^{-1}$, with $\tau_{sat}=\SI{1}{\milli\second}$.}\label{fig:results}
\end{figure*}
The central result of this paper is presented in \cref{fig:results}. Figures~\labelcref{fig:results}(a,b) show the measured cross-PSDs and Figures~\labelcref{fig:results}(c,d) show the extracted apparent quasiparticle lifetime and fluctuation level from Lorentzians fits to these spectra (\cref{eq:Sn}). For high temperatures (\SI{\geq 300}{\milli\kelvin}), we observe a higher level and $\tau_{qp}^*$ for the membrane resonator compared to the substrate resonator. This is expected from \cref{eq:tqpstar,eq:Sn}, as the membrane effectively increases $\tau_{esc}$. The dashed lines in \cref{fig:results}(c,d) are fits to \cref{eq:tqp,eq:tqpstar} for temperatures \SI{\geq 300}{\milli\kelvin}, with $\tau_{esc}$ as only free parameter, resulting in $\tau_{esc}=\SI{5.6\pm 0.4}{\nano\second}$ and \SI{0.09\pm0.02}{\nano\second} for the membrane and substrate resonator respectively. These intervals only indicate uncertainties from the fitting procedure. Equivalently, the phonon trapping factors (\cref{eq:tqpstar}) are 21 and 1.3 from which we would expect a \SI{12}{\decibel} higher fluctuation level for the membrane resonator. This is indeed observed in \cref{fig:results}(d). The dashed lines give the expected level from \cref{eq:Sn}, where the fitted $\tau_{esc}$ is used. A detailed analysis of the effects of the membrane on phonon statistics and energy resolution will be published elsewhere \cite{deVisser2021a}.\\
The fluctuation levels expected from \cref{eq:Sn} (dashed lines in \cref{fig:results}(d)) are constant with changing temperature. In sharp contrast, we here observe a strongly decreasing level when lowering the temperature. At temperatures \SI{<190}{\milli\kelvin}, no Lorentzian spectrum (\cref{eq:Sn}) can be identified. The individual amplitude and phase PSDs show this behavior as well \footnotemark[1]. As the responsivity factor in \cref{eq:Sn_exp} is constant in this temperature range \footnotemark[1], this result should be interpreted as a strong reduction of $S_{N_{qp}}$.\\
\begin{figure}
	\includegraphics[width=\linewidth]{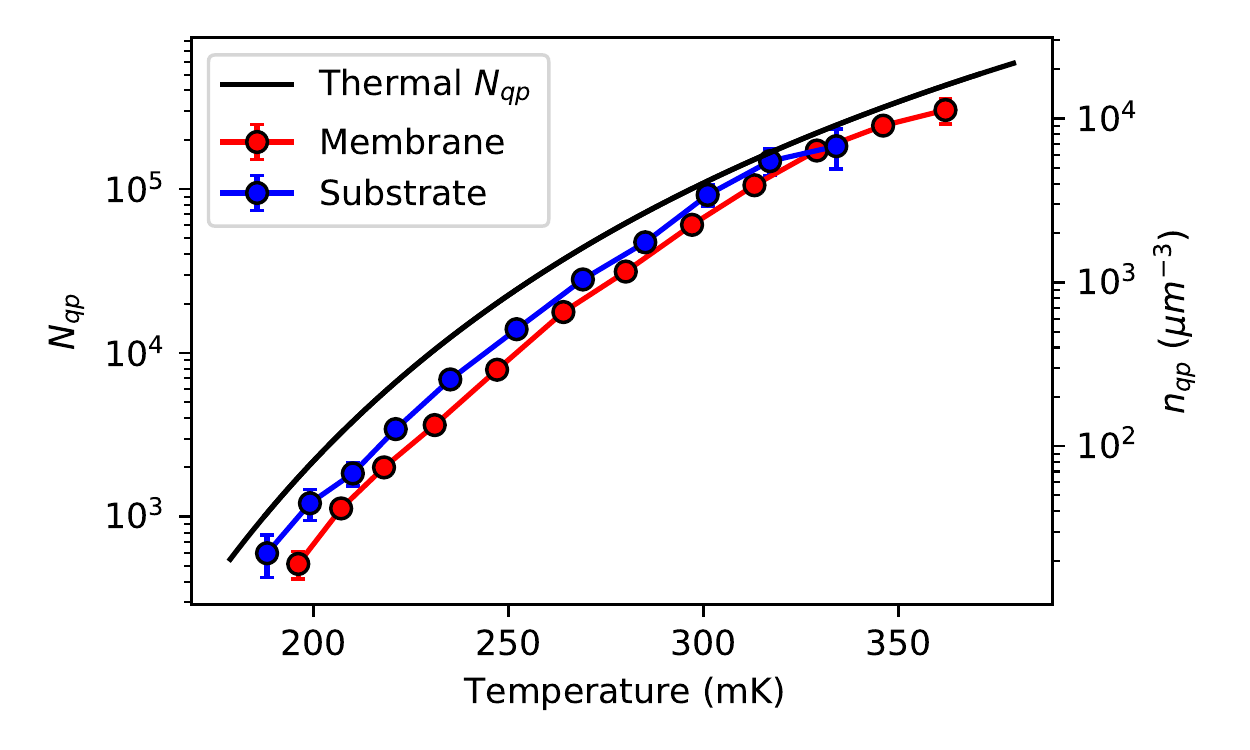}
	\caption{Number of quasiparticles and quasiparticle density (right axis) within the Al volume, calculated from the Lorentzian fits in \cref{fig:results}(c,d). Error bars indicate propagated fitting uncertainties. The solid black line is from \cref{eq:Nqp}.}\label{fig:Nqp}
\end{figure}
We calculate $N_{qp}$ from $S_{N_{qp}}$ by dividing the level by $4\tau_{qp}^*$, which is shown in \cref{fig:Nqp}. The number of quasiparticles follows a thermal dependence for both the substrate and membrane resonator. Therefore, we conclude that the lifetime saturation is not caused by excess quasiparticles, in contrast to the observations in Ref. \cite{deVisser2011}. On top of that, the lifetimes observed for membrane and substrate resonators, saturate at a similar level despite the 16 times stronger phonon trapping in the membrane. Therefore the lifetime saturation must originate from an effect that directly interacts with the quasiparticle system.\\
The design of the NbTiN-Al hybrid resonators enables read-out low microwave powers ($P_{read}$), which minimizes the creation of excess quasiparticles. When we increase $P_{read}$, we eventually observe excess quasiparticles without a level reduction, equivalent to a higher, $P_{read}$-induced, effective temperature. The results of Refs. \cite{deVisser2011,deVisser2012} are thus recovered in the high $P_{read}$-regime. See the Supplemental Material for more details \footnotemark[1].\\
To verify that the lifetime saturation without creation of excess quasiparticles results in a level reduction, we calculated the lifetimes and fluctuation levels with \cref{eq:Sn}, but with an alteration to \cref{eq:tqpstar}: 
\begin{equation}\label{eq:tqpstarsat}
	\frac{1}{\tau_{qp}^*}=\frac{2}{\tau_{qp}(1+\tau_{esc}/\tau_{pb})}+\frac{1}{\tau_{sat}},
\end{equation}
with $\tau_{sat}$ the saturation lifetime. For both membrane and substrate, $\tau_{sat}$ is set to \SI{1}{\milli\second}, based on $\tau_{qp}^*$ from the membrane resonator. The results are plotted in \cref{fig:results}(c,d) as the dotted lines, which follow the measured $\tau_{qp}^*$ very well and shows a decrease in fluctuation level, for both resonators. The deviations of level in \cref{fig:results}(d), are likely caused by the responsivity measurement method \footnotemark[1].\\

A possible cause of a lifetime saturation without excess quasiparticles is \textit{quasiparticle trapping}. From theory, it is known that magnetic \cite{Abrikosov1959,*Abrikosov1959a,Muller-Hartmann1971,Yu1965,*Shiba1968,*Rusinov1969,Fominov2011} and non-magnetic \cite{Kaiser1970,Ghosal2001,Bespalov2016} impurities can cause subgap electronic states \cite{Bespalov2019} and disorder, which can result in local variations of $\Delta$ \cite{Larkin1972,Feigelman2012}. Also thickness variations \cite{Chubov1969} and unpaired surface spins \cite{Faoro2008} due to native oxide \cite{Kumar2016} may induce $\Delta$-variations. Quasiparticles can be trapped at these suppressed $\Delta$-regions \cite{Bespalov2016a,Meyer2020} by inelastic scattering, resulting in a background number of trapped quasiparticles, $N_t$. At low temperatures, $N_{qp}$ (\cref{eq:Nqp}) inevitably becomes comparable to $N_t$, leading to dominating quasiparticle trapping behavior even for low impurity concentrations, weak disorder or other effects that lead to small variations in $\Delta$ \cite{Bespalov2016a,Kozorezov2008,Kozorezov2009}.\\
At that point, the quasiparticle lifetime is no longer limited by free quasiparticle recombination, but by \textit{on-trap} recombination events, where a free quasiparticle recombines with a trapped one. Trapping itself does not change the number of Cooper-pairs and therefore is not observable in our experiments. This is in contrast to experiments where the free quasiparticle density near a junction is measured, in which case trapping dominates the low temperature behavior \cite{Kozorezov2001,Kozorezov2008a}. The trapping states we conjecture are dissipative, as the amplitude and phase PSDs show the same temperature behavior \footnotemark[1], in contrast to what is observed for disordered TiN \cite{Gao2012,Bueno2014}.\\
Analogous to \cref{eq:tqpstar,eq:tqp}, the saturation lifetime due to on-trap recombination can be written, $\tau_{sat} = V/2 R_t N_t$, with $R_t$ the on-trap recombination constant \cite{Kozorezov2008}. The phonon that is emitted during an on-trap recombination event has an energy $\Omega<2\Delta$ and is therefore unable to break a Cooper-pair into two free quasiparticles. \textit{On-trap} pair-breaking (a subgap phonon breaking a Cooper-pair into a trapped and free quasiparticle) is far less likely because the density of trapping states is assumed to be small \cite{Kozorezov2008}. Therefore, $\tau_{sat}$ is independent of phonon trapping, which is consistent with the observation that the substrate and membrane resonator show the same $\tau_{sat}$.\\
To investigate where quasiparticle traps could be located in our system, we conducted the same experiment with different geometries of the Al strip. Width (\SIrange{0.6}{1.5}{\micro\meter}) and length (\SIrange{0.12}{1.4}{\milli\meter}) variations did not affect the saturation lifetime, which implies the traps are not located at the NbTiN-Al interface or the sides of the Al central line. $\tau_{sat}$ increases from \SIrange{.7}{3.1}{\milli\second} with increasing the film thickness from \SIrange{25}{150}{\nano\meter}, respectively, and the reduction in fluctuation level is not observed for the thickest film. The experimental data of these geometry variations is presented in the Supplemental Material \footnotemark[1]. This suggests that the trap density ($N_t/V$) decreases for thicker films and the location of the traps is either at the Al-substrate interface or the top surface. The top surface might contain unpaired surface spins from native Al oxide \cite{Kumar2016,Barends2009a}. Further progress in understanding the trapping mechanism will require identification of the microscopic origin of the traps.\\

We here analysed the lifetime saturation phenomenologically using \cref{eq:tqpstarsat}. A more detailed model of the fluctuations comprises multivariable rate equations for quasiparticles and phonons, including quasiparticle trapping and on-trap recombination terms \cite{Kozorezov2008,Wilson2004}. Only non-equilibrium experiments \cite{Kozorezov2008a,Grunhaupt2018,Wang2014} were conducted previously ($\delta N_{qp}\gg N_{qp}$), which allows to neglect the generation terms and feedback from the phonon system, in contrast to our near-equilibrium ($\delta N_{qp} \ll N_{qp}$) experiment. We have studied several such models \cite{deRooij2020}, but without satisfactory results due to the number of (unknown) input parameters. However, we can constrain such a model to follow our experimental data.\\

Quasiparticle trapping could decrease the low-temperature responsivity of superconducting tunnel junction detectors \cite{Kozorezov2001}, cause long-lived excitations in granular Al resonators \cite{Grunhaupt2018} and lead to anomalous electrodynamic response of disordered TiN resonators \cite{Gao2012,Bueno2014}. However, for single-photon detection using MKIDs\cite{Day2003,Zmuidzinas2012} , the quasiparticle trapping effects can be beneficial. The fluctuation level reduction results in a higher signal-to-noise ratio when generation-recombination noise is the dominant noise source and the responsivity stays the same, as we observe \footnotemark[1]\cite{deVisser2021a}. This means, somewhat counter-intuitively, that single-photon detector performance can be improved at low temperatures, by introducing quasiparticle traps, compared to a quasiparticle number saturation \cite{deVisser2012}.\\

In conclusion, at high temperatures (\SI{>300}{\milli\kelvin}), quasiparticle lifetimes in Al superconducting resonators are well described by simple recombination in thermal equilibrium \cite{Kaplan1976}. At low temperatures, the lifetime is limited to \SI{<1}{\milli\second}, independent of phonon trapping, while the quasiparticle number exponentially decreases with temperature. This results in a strong reduction of quasiparticle fluctuations. Quasiparticle trapping with subsequent on-trap recombination is consistent with our observations. Variations in the Al strip geometry (width, length and thickness) showed that the traps are likely located at top or bottom surfaces.\\

All presented data and used analysis scripts are available at: \url{https://doi.org/10.5281/zenodo.4580356} .\\

\begin{acknowledgments}
	We acknowledge useful discussions with Yaroslav Blanter and Karwan Rostem and experimental assistance from Juan Bueno and Sven Visser.	P.J. de V. was supported by the Netherlands Organisation for Scientific Research NWO (Veni Grant No. 639.041.750 and Projectruimte 680-91-127). J.J.A.B. was supported by the European Research Council ERC (Consolidator Grant No. 648135 MOSAIC). 
\end{acknowledgments}

\bibliography{GRNoise}

\begin{thebibliography}{64}%
\makeatletter
\providecommand \@ifxundefined [1]{%
 \@ifx{#1\undefined}
}%
\providecommand \@ifnum [1]{%
 \ifnum #1\expandafter \@firstoftwo
 \else \expandafter \@secondoftwo
 \fi
}%
\providecommand \@ifx [1]{%
 \ifx #1\expandafter \@firstoftwo
 \else \expandafter \@secondoftwo
 \fi
}%
\providecommand \natexlab [1]{#1}%
\providecommand \enquote  [1]{``#1''}%
\providecommand \bibnamefont  [1]{#1}%
\providecommand \bibfnamefont [1]{#1}%
\providecommand \citenamefont [1]{#1}%
\providecommand \href@noop [0]{\@secondoftwo}%
\providecommand \href [0]{\begingroup \@sanitize@url \@href}%
\providecommand \@href[1]{\@@startlink{#1}\@@href}%
\providecommand \@@href[1]{\endgroup#1\@@endlink}%
\providecommand \@sanitize@url [0]{\catcode `\\12\catcode `\$12\catcode
  `\&12\catcode `\#12\catcode `\^12\catcode `\_12\catcode `\%12\relax}%
\providecommand \@@startlink[1]{}%
\providecommand \@@endlink[0]{}%
\providecommand \url  [0]{\begingroup\@sanitize@url \@url }%
\providecommand \@url [1]{\endgroup\@href {#1}{\urlprefix }}%
\providecommand \urlprefix  [0]{URL }%
\providecommand \Eprint [0]{\href }%
\providecommand \doibase [0]{https://doi.org/}%
\providecommand \selectlanguage [0]{\@gobble}%
\providecommand \bibinfo  [0]{\@secondoftwo}%
\providecommand \bibfield  [0]{\@secondoftwo}%
\providecommand \translation [1]{[#1]}%
\providecommand \BibitemOpen [0]{}%
\providecommand \bibitemStop [0]{}%
\providecommand \bibitemNoStop [0]{.\EOS\space}%
\providecommand \EOS [0]{\spacefactor3000\relax}%
\providecommand \BibitemShut  [1]{\csname bibitem#1\endcsname}%
\let\auto@bib@innerbib\@empty
\bibitem [{\citenamefont {Wilson}\ and\ \citenamefont
  {Prober}(2004)}]{Wilson2004}%
  \BibitemOpen
  \bibfield  {author} {\bibinfo {author} {\bibfnamefont {C.~M.}\ \bibnamefont
  {Wilson}}\ and\ \bibinfo {author} {\bibfnamefont {D.~E.}\ \bibnamefont
  {Prober}},\ }\bibfield  {title} {\bibinfo {title} {Quasiparticle number
  fluctuations in superconductors},\ }\href
  {https://doi.org/10.1103/PhysRevB.69.094524} {\bibfield  {journal} {\bibinfo
  {journal} {Phys. Rev. B}\ }\textbf {\bibinfo {volume} {69}},\ \bibinfo
  {pages} {094524} (\bibinfo {year} {2004})}\BibitemShut {NoStop}%
\bibitem [{\citenamefont {{de Visser}}\ \emph {et~al.}(2011)\citenamefont {{de
  Visser}}, \citenamefont {Baselmans}, \citenamefont {Diener}, \citenamefont
  {Yates}, \citenamefont {Endo},\ and\ \citenamefont
  {Klapwijk}}]{deVisser2011}%
  \BibitemOpen
  \bibfield  {author} {\bibinfo {author} {\bibfnamefont {P.~J.}\ \bibnamefont
  {{de Visser}}}, \bibinfo {author} {\bibfnamefont {J.~J.~A.}\ \bibnamefont
  {Baselmans}}, \bibinfo {author} {\bibfnamefont {P.}~\bibnamefont {Diener}},
  \bibinfo {author} {\bibfnamefont {S.~J.~C.}\ \bibnamefont {Yates}}, \bibinfo
  {author} {\bibfnamefont {A.}~\bibnamefont {Endo}},\ and\ \bibinfo {author}
  {\bibfnamefont {T.~M.}\ \bibnamefont {Klapwijk}},\ }\bibfield  {title}
  {\bibinfo {title} {Number {{Fluctuations}} of {{Sparse Quasiparticles}} in a
  {{Superconductor}}},\ }\href {https://doi.org/10.1103/PhysRevLett.106.167004}
  {\bibfield  {journal} {\bibinfo  {journal} {Phys. Rev. Lett.}\ }\textbf
  {\bibinfo {volume} {106}},\ \bibinfo {pages} {167004} (\bibinfo {year}
  {2011})}\BibitemShut {NoStop}%
\bibitem [{\citenamefont {Kaplan}\ \emph {et~al.}(1976)\citenamefont {Kaplan},
  \citenamefont {Chi}, \citenamefont {Langenberg}, \citenamefont {Chang},
  \citenamefont {Jafarey},\ and\ \citenamefont {Scalapino}}]{Kaplan1976}%
  \BibitemOpen
  \bibfield  {author} {\bibinfo {author} {\bibfnamefont {S.~B.}\ \bibnamefont
  {Kaplan}}, \bibinfo {author} {\bibfnamefont {C.~C.}\ \bibnamefont {Chi}},
  \bibinfo {author} {\bibfnamefont {D.~N.}\ \bibnamefont {Langenberg}},
  \bibinfo {author} {\bibfnamefont {J.~J.}\ \bibnamefont {Chang}}, \bibinfo
  {author} {\bibfnamefont {S.}~\bibnamefont {Jafarey}},\ and\ \bibinfo {author}
  {\bibfnamefont {D.~J.}\ \bibnamefont {Scalapino}},\ }\bibfield  {title}
  {\bibinfo {title} {Quasiparticle and phonon lifetimes in superconductors},\
  }\href {https://doi.org/10.1103/PhysRevB.14.4854} {\bibfield  {journal}
  {\bibinfo  {journal} {Phys. Rev. B}\ }\textbf {\bibinfo {volume} {14}},\
  \bibinfo {pages} {4854} (\bibinfo {year} {1976})}\BibitemShut {NoStop}%
\bibitem [{\citenamefont {Martinis}\ \emph {et~al.}(2009)\citenamefont
  {Martinis}, \citenamefont {Ansmann},\ and\ \citenamefont
  {Aumentado}}]{Martinis2009}%
  \BibitemOpen
  \bibfield  {author} {\bibinfo {author} {\bibfnamefont {J.~M.}\ \bibnamefont
  {Martinis}}, \bibinfo {author} {\bibfnamefont {M.}~\bibnamefont {Ansmann}},\
  and\ \bibinfo {author} {\bibfnamefont {J.}~\bibnamefont {Aumentado}},\
  }\bibfield  {title} {\bibinfo {title} {Energy {{Decay}} in {{Superconducting
  Josephson}}-{{Junction Qubits}} from {{Nonequilibrium Quasiparticle
  Excitations}}},\ }\href {https://doi.org/10.1103/PhysRevLett.103.097002}
  {\bibfield  {journal} {\bibinfo  {journal} {Phys. Rev. Lett.}\ }\textbf
  {\bibinfo {volume} {103}},\ \bibinfo {pages} {097002} (\bibinfo {year}
  {2009})}\BibitemShut {NoStop}%
\bibitem [{\citenamefont {Rainis}\ and\ \citenamefont
  {Loss}(2012)}]{Rainis2012}%
  \BibitemOpen
  \bibfield  {author} {\bibinfo {author} {\bibfnamefont {D.}~\bibnamefont
  {Rainis}}\ and\ \bibinfo {author} {\bibfnamefont {D.}~\bibnamefont {Loss}},\
  }\bibfield  {title} {\bibinfo {title} {Majorana qubit decoherence by
  quasiparticle poisoning},\ }\href
  {https://doi.org/10.1103/PhysRevB.85.174533} {\bibfield  {journal} {\bibinfo
  {journal} {Phys. Rev. B}\ }\textbf {\bibinfo {volume} {85}},\ \bibinfo
  {pages} {174533} (\bibinfo {year} {2012})}\BibitemShut {NoStop}%
\bibitem [{\citenamefont {Barends}\ \emph {et~al.}(2011)\citenamefont
  {Barends}, \citenamefont {Wenner}, \citenamefont {Lenander}, \citenamefont
  {Chen}, \citenamefont {Bialczak}, \citenamefont {Kelly}, \citenamefont
  {Lucero}, \citenamefont {O'Malley}, \citenamefont {Mariantoni}, \citenamefont
  {Sank}, \citenamefont {Wang}, \citenamefont {White}, \citenamefont {Yin},
  \citenamefont {Zhao}, \citenamefont {Cleland}, \citenamefont {Martinis},\
  and\ \citenamefont {Baselmans}}]{Barends2011}%
  \BibitemOpen
  \bibfield  {author} {\bibinfo {author} {\bibfnamefont {R.}~\bibnamefont
  {Barends}}, \bibinfo {author} {\bibfnamefont {J.}~\bibnamefont {Wenner}},
  \bibinfo {author} {\bibfnamefont {M.}~\bibnamefont {Lenander}}, \bibinfo
  {author} {\bibfnamefont {Y.}~\bibnamefont {Chen}}, \bibinfo {author}
  {\bibfnamefont {R.~C.}\ \bibnamefont {Bialczak}}, \bibinfo {author}
  {\bibfnamefont {J.}~\bibnamefont {Kelly}}, \bibinfo {author} {\bibfnamefont
  {E.}~\bibnamefont {Lucero}}, \bibinfo {author} {\bibfnamefont
  {P.}~\bibnamefont {O'Malley}}, \bibinfo {author} {\bibfnamefont
  {M.}~\bibnamefont {Mariantoni}}, \bibinfo {author} {\bibfnamefont
  {D.}~\bibnamefont {Sank}}, \bibinfo {author} {\bibfnamefont {H.}~\bibnamefont
  {Wang}}, \bibinfo {author} {\bibfnamefont {T.~C.}\ \bibnamefont {White}},
  \bibinfo {author} {\bibfnamefont {Y.}~\bibnamefont {Yin}}, \bibinfo {author}
  {\bibfnamefont {J.}~\bibnamefont {Zhao}}, \bibinfo {author} {\bibfnamefont
  {A.~N.}\ \bibnamefont {Cleland}}, \bibinfo {author} {\bibfnamefont {J.~M.}\
  \bibnamefont {Martinis}},\ and\ \bibinfo {author} {\bibfnamefont {J.~J.~A.}\
  \bibnamefont {Baselmans}},\ }\bibfield  {title} {\bibinfo {title} {Minimizing
  quasiparticle generation from stray infrared light in superconducting quantum
  circuits},\ }\href {https://doi.org/10.1063/1.3638063} {\bibfield  {journal}
  {\bibinfo  {journal} {Appl. Phys. Lett.}\ }\textbf {\bibinfo {volume} {99}},\
  \bibinfo {pages} {113507} (\bibinfo {year} {2011})}\BibitemShut {NoStop}%
\bibitem [{\citenamefont {Karatsu}\ \emph {et~al.}(2019)\citenamefont
  {Karatsu}, \citenamefont {Endo}, \citenamefont {Bueno}, \citenamefont {{de
  Visser}}, \citenamefont {Barends}, \citenamefont {Thoen}, \citenamefont
  {Murugesan}, \citenamefont {Tomita},\ and\ \citenamefont
  {Baselmans}}]{Karatsu2019}%
  \BibitemOpen
  \bibfield  {author} {\bibinfo {author} {\bibfnamefont {K.}~\bibnamefont
  {Karatsu}}, \bibinfo {author} {\bibfnamefont {A.}~\bibnamefont {Endo}},
  \bibinfo {author} {\bibfnamefont {J.}~\bibnamefont {Bueno}}, \bibinfo
  {author} {\bibfnamefont {P.~J.}\ \bibnamefont {{de Visser}}}, \bibinfo
  {author} {\bibfnamefont {R.}~\bibnamefont {Barends}}, \bibinfo {author}
  {\bibfnamefont {D.~J.}\ \bibnamefont {Thoen}}, \bibinfo {author}
  {\bibfnamefont {V.}~\bibnamefont {Murugesan}}, \bibinfo {author}
  {\bibfnamefont {N.}~\bibnamefont {Tomita}},\ and\ \bibinfo {author}
  {\bibfnamefont {J.~J.~A.}\ \bibnamefont {Baselmans}},\ }\bibfield  {title}
  {\bibinfo {title} {Mitigation of cosmic ray effect on microwave kinetic
  inductance detector arrays},\ }\href {https://doi.org/10.1063/1.5052419}
  {\bibfield  {journal} {\bibinfo  {journal} {Appl. Phys. Lett.}\ }\textbf
  {\bibinfo {volume} {114}},\ \bibinfo {pages} {032601} (\bibinfo {year}
  {2019})}\BibitemShut {NoStop}%
\bibitem [{\citenamefont {Cardani}\ \emph {et~al.}(2021)\citenamefont
  {Cardani}, \citenamefont {Valenti}, \citenamefont {Casali}, \citenamefont
  {Catelani}, \citenamefont {Charpentier}, \citenamefont {Clemenza},
  \citenamefont {Colantoni}, \citenamefont {Cruciani}, \citenamefont
  {D'Imperio}, \citenamefont {Gironi}, \citenamefont {Gr{\"u}nhaupt},
  \citenamefont {Gusenkova}, \citenamefont {Henriques}, \citenamefont {Lagoin},
  \citenamefont {Martinez}, \citenamefont {Pettinari}, \citenamefont {Rusconi},
  \citenamefont {Sander}, \citenamefont {Tomei}, \citenamefont {Ustinov},
  \citenamefont {Weber}, \citenamefont {Wernsdorfer}, \citenamefont {Vignati},
  \citenamefont {Pirro},\ and\ \citenamefont {Pop}}]{Cardani2021}%
  \BibitemOpen
  \bibfield  {author} {\bibinfo {author} {\bibfnamefont {L.}~\bibnamefont
  {Cardani}}, \bibinfo {author} {\bibfnamefont {F.}~\bibnamefont {Valenti}},
  \bibinfo {author} {\bibfnamefont {N.}~\bibnamefont {Casali}}, \bibinfo
  {author} {\bibfnamefont {G.}~\bibnamefont {Catelani}}, \bibinfo {author}
  {\bibfnamefont {T.}~\bibnamefont {Charpentier}}, \bibinfo {author}
  {\bibfnamefont {M.}~\bibnamefont {Clemenza}}, \bibinfo {author}
  {\bibfnamefont {I.}~\bibnamefont {Colantoni}}, \bibinfo {author}
  {\bibfnamefont {A.}~\bibnamefont {Cruciani}}, \bibinfo {author}
  {\bibfnamefont {G.}~\bibnamefont {D'Imperio}}, \bibinfo {author}
  {\bibfnamefont {L.}~\bibnamefont {Gironi}}, \bibinfo {author} {\bibfnamefont
  {L.}~\bibnamefont {Gr{\"u}nhaupt}}, \bibinfo {author} {\bibfnamefont
  {D.}~\bibnamefont {Gusenkova}}, \bibinfo {author} {\bibfnamefont
  {F.}~\bibnamefont {Henriques}}, \bibinfo {author} {\bibfnamefont
  {M.}~\bibnamefont {Lagoin}}, \bibinfo {author} {\bibfnamefont
  {M.}~\bibnamefont {Martinez}}, \bibinfo {author} {\bibfnamefont
  {G.}~\bibnamefont {Pettinari}}, \bibinfo {author} {\bibfnamefont
  {C.}~\bibnamefont {Rusconi}}, \bibinfo {author} {\bibfnamefont
  {O.}~\bibnamefont {Sander}}, \bibinfo {author} {\bibfnamefont
  {C.}~\bibnamefont {Tomei}}, \bibinfo {author} {\bibfnamefont {A.~V.}\
  \bibnamefont {Ustinov}}, \bibinfo {author} {\bibfnamefont {M.}~\bibnamefont
  {Weber}}, \bibinfo {author} {\bibfnamefont {W.}~\bibnamefont {Wernsdorfer}},
  \bibinfo {author} {\bibfnamefont {M.}~\bibnamefont {Vignati}}, \bibinfo
  {author} {\bibfnamefont {S.}~\bibnamefont {Pirro}},\ and\ \bibinfo {author}
  {\bibfnamefont {I.~M.}\ \bibnamefont {Pop}},\ }\bibfield  {title} {\bibinfo
  {title} {Reducing the impact of radioactivity on quantum circuits in a
  deep-underground facility},\ }\href
  {https://doi.org/10.1038/s41467-021-23032-z} {\bibfield  {journal} {\bibinfo
  {journal} {Nature Communications}\ }\textbf {\bibinfo {volume} {12}},\
  \bibinfo {pages} {2733} (\bibinfo {year} {2021})}\BibitemShut {NoStop}%
\bibitem [{\citenamefont {{de Visser}}\ \emph
  {et~al.}(2014{\natexlab{a}})\citenamefont {{de Visser}}, \citenamefont
  {Goldie}, \citenamefont {Diener}, \citenamefont {Withington}, \citenamefont
  {Baselmans},\ and\ \citenamefont {Klapwijk}}]{deVisser2014a}%
  \BibitemOpen
  \bibfield  {author} {\bibinfo {author} {\bibfnamefont {P.~J.}\ \bibnamefont
  {{de Visser}}}, \bibinfo {author} {\bibfnamefont {D.~J.}\ \bibnamefont
  {Goldie}}, \bibinfo {author} {\bibfnamefont {P.}~\bibnamefont {Diener}},
  \bibinfo {author} {\bibfnamefont {S.}~\bibnamefont {Withington}}, \bibinfo
  {author} {\bibfnamefont {J.~J.~A.}\ \bibnamefont {Baselmans}},\ and\ \bibinfo
  {author} {\bibfnamefont {T.~M.}\ \bibnamefont {Klapwijk}},\ }\bibfield
  {title} {\bibinfo {title} {Evidence of a {{Nonequilibrium Distribution}} of
  {{Quasiparticles}} in the {{Microwave Response}} of a {{Superconducting
  Aluminum Resonator}}},\ }\href
  {https://doi.org/10.1103/PhysRevLett.112.047004} {\bibfield  {journal}
  {\bibinfo  {journal} {Phys. Rev. Lett.}\ }\textbf {\bibinfo {volume} {112}},\
  \bibinfo {pages} {047004} (\bibinfo {year} {2014}{\natexlab{a}})}\BibitemShut
  {NoStop}%
\bibitem [{\citenamefont {Barends}\ \emph {et~al.}(2009)\citenamefont
  {Barends}, \citenamefont {{van Vliet}}, \citenamefont {Baselmans},
  \citenamefont {Yates}, \citenamefont {Gao},\ and\ \citenamefont
  {Klapwijk}}]{Barends2009a}%
  \BibitemOpen
  \bibfield  {author} {\bibinfo {author} {\bibfnamefont {R.}~\bibnamefont
  {Barends}}, \bibinfo {author} {\bibfnamefont {S.}~\bibnamefont {{van
  Vliet}}}, \bibinfo {author} {\bibfnamefont {J.~J.~A.}\ \bibnamefont
  {Baselmans}}, \bibinfo {author} {\bibfnamefont {S.~J.~C.}\ \bibnamefont
  {Yates}}, \bibinfo {author} {\bibfnamefont {J.~R.}\ \bibnamefont {Gao}},\
  and\ \bibinfo {author} {\bibfnamefont {T.~M.}\ \bibnamefont {Klapwijk}},\
  }\bibfield  {title} {\bibinfo {title} {Enhancement of quasiparticle
  recombination in {{Ta}} and {{Al}} superconductors by implantation of
  magnetic and nonmagnetic atoms},\ }\href
  {https://doi.org/10.1103/PhysRevB.79.020509} {\bibfield  {journal} {\bibinfo
  {journal} {Phys. Rev. B}\ }\textbf {\bibinfo {volume} {79}},\ \bibinfo
  {pages} {020509(R)} (\bibinfo {year} {2009})}\BibitemShut {NoStop}%
\bibitem [{\citenamefont {Gao}\ \emph {et~al.}(2012)\citenamefont {Gao},
  \citenamefont {Vissers}, \citenamefont {Sandberg}, \citenamefont {{da
  Silva}}, \citenamefont {Nam}, \citenamefont {Pappas}, \citenamefont {Wisbey},
  \citenamefont {Langman}, \citenamefont {Meeker}, \citenamefont {Mazin},
  \citenamefont {Leduc}, \citenamefont {Zmuidzinas},\ and\ \citenamefont
  {Irwin}}]{Gao2012}%
  \BibitemOpen
  \bibfield  {author} {\bibinfo {author} {\bibfnamefont {J.}~\bibnamefont
  {Gao}}, \bibinfo {author} {\bibfnamefont {M.~R.}\ \bibnamefont {Vissers}},
  \bibinfo {author} {\bibfnamefont {M.~O.}\ \bibnamefont {Sandberg}}, \bibinfo
  {author} {\bibfnamefont {F.~C.~S.}\ \bibnamefont {{da Silva}}}, \bibinfo
  {author} {\bibfnamefont {S.~W.}\ \bibnamefont {Nam}}, \bibinfo {author}
  {\bibfnamefont {D.~P.}\ \bibnamefont {Pappas}}, \bibinfo {author}
  {\bibfnamefont {D.~S.}\ \bibnamefont {Wisbey}}, \bibinfo {author}
  {\bibfnamefont {E.~C.}\ \bibnamefont {Langman}}, \bibinfo {author}
  {\bibfnamefont {S.~R.}\ \bibnamefont {Meeker}}, \bibinfo {author}
  {\bibfnamefont {B.~A.}\ \bibnamefont {Mazin}}, \bibinfo {author}
  {\bibfnamefont {H.~G.}\ \bibnamefont {Leduc}}, \bibinfo {author}
  {\bibfnamefont {J.}~\bibnamefont {Zmuidzinas}},\ and\ \bibinfo {author}
  {\bibfnamefont {K.~D.}\ \bibnamefont {Irwin}},\ }\bibfield  {title} {\bibinfo
  {title} {A titanium-nitride near-infrared kinetic inductance photon-counting
  detector and its anomalous electrodynamics},\ }\href
  {https://doi.org/10.1063/1.4756916} {\bibfield  {journal} {\bibinfo
  {journal} {Appl. Phys. Lett.}\ }\textbf {\bibinfo {volume} {101}},\ \bibinfo
  {pages} {142602} (\bibinfo {year} {2012})}\BibitemShut {NoStop}%
\bibitem [{\citenamefont {Barends}\ \emph {et~al.}(2008)\citenamefont
  {Barends}, \citenamefont {Baselmans}, \citenamefont {Yates}, \citenamefont
  {Gao}, \citenamefont {Hovenier},\ and\ \citenamefont
  {Klapwijk}}]{Barends2008}%
  \BibitemOpen
  \bibfield  {author} {\bibinfo {author} {\bibfnamefont {R.}~\bibnamefont
  {Barends}}, \bibinfo {author} {\bibfnamefont {J.~J.~A.}\ \bibnamefont
  {Baselmans}}, \bibinfo {author} {\bibfnamefont {S.~J.~C.}\ \bibnamefont
  {Yates}}, \bibinfo {author} {\bibfnamefont {J.~R.}\ \bibnamefont {Gao}},
  \bibinfo {author} {\bibfnamefont {J.~N.}\ \bibnamefont {Hovenier}},\ and\
  \bibinfo {author} {\bibfnamefont {T.~M.}\ \bibnamefont {Klapwijk}},\
  }\bibfield  {title} {\bibinfo {title} {Quasiparticle {{Relaxation}} in
  {{Optically Excited High}}-{{Q Superconducting Resonators}}},\ }\href
  {https://doi.org/10.1103/PhysRevLett.100.257002} {\bibfield  {journal}
  {\bibinfo  {journal} {Phys. Rev. Lett.}\ }\textbf {\bibinfo {volume} {100}},\
  \bibinfo {pages} {257002} (\bibinfo {year} {2008})}\BibitemShut {NoStop}%
\bibitem [{\citenamefont {Fyhrie}\ \emph {et~al.}(2020)\citenamefont {Fyhrie},
  \citenamefont {Day}, \citenamefont {Glenn}, \citenamefont {Leduc},
  \citenamefont {McKenney}, \citenamefont {Perido},\ and\ \citenamefont
  {Zmuidzinas}}]{Fyhrie2020}%
  \BibitemOpen
  \bibfield  {author} {\bibinfo {author} {\bibfnamefont {A.}~\bibnamefont
  {Fyhrie}}, \bibinfo {author} {\bibfnamefont {P.}~\bibnamefont {Day}},
  \bibinfo {author} {\bibfnamefont {J.}~\bibnamefont {Glenn}}, \bibinfo
  {author} {\bibfnamefont {H.}~\bibnamefont {Leduc}}, \bibinfo {author}
  {\bibfnamefont {C.}~\bibnamefont {McKenney}}, \bibinfo {author}
  {\bibfnamefont {J.}~\bibnamefont {Perido}},\ and\ \bibinfo {author}
  {\bibfnamefont {J.}~\bibnamefont {Zmuidzinas}},\ }\bibfield  {title}
  {\bibinfo {title} {Decay {{Times}} of {{Optical Pulses}} for {{Aluminum CPW
  KIDs}}},\ }\href {https://doi.org/10.1007/s10909-020-02377-7} {\bibfield
  {journal} {\bibinfo  {journal} {J Low Temp Phys}\ }\textbf {\bibinfo {volume}
  {199}},\ \bibinfo {pages} {688} (\bibinfo {year} {2020})}\BibitemShut
  {NoStop}%
\bibitem [{\citenamefont {{de Visser}}\ \emph {et~al.}(2012)\citenamefont {{de
  Visser}}, \citenamefont {Baselmans}, \citenamefont {Yates}, \citenamefont
  {Diener}, \citenamefont {Endo},\ and\ \citenamefont
  {Klapwijk}}]{deVisser2012}%
  \BibitemOpen
  \bibfield  {author} {\bibinfo {author} {\bibfnamefont {P.~J.}\ \bibnamefont
  {{de Visser}}}, \bibinfo {author} {\bibfnamefont {J.~J.~A.}\ \bibnamefont
  {Baselmans}}, \bibinfo {author} {\bibfnamefont {S.~J.~C.}\ \bibnamefont
  {Yates}}, \bibinfo {author} {\bibfnamefont {P.}~\bibnamefont {Diener}},
  \bibinfo {author} {\bibfnamefont {A.}~\bibnamefont {Endo}},\ and\ \bibinfo
  {author} {\bibfnamefont {T.~M.}\ \bibnamefont {Klapwijk}},\ }\bibfield
  {title} {\bibinfo {title} {Microwave-induced excess quasiparticles in
  superconducting resonators measured through correlated conductivity
  fluctuations},\ }\href {https://doi.org/10.1063/1.4704151} {\bibfield
  {journal} {\bibinfo  {journal} {Appl. Phys. Lett.}\ }\textbf {\bibinfo
  {volume} {100}},\ \bibinfo {pages} {162601} (\bibinfo {year}
  {2012})}\BibitemShut {NoStop}%
\bibitem [{\citenamefont {Riwar}\ \emph {et~al.}(2016)\citenamefont {Riwar},
  \citenamefont {Hosseinkhani}, \citenamefont {Burkhart}, \citenamefont {Gao},
  \citenamefont {Schoelkopf}, \citenamefont {Glazman},\ and\ \citenamefont
  {Catelani}}]{Riwar2016}%
  \BibitemOpen
  \bibfield  {author} {\bibinfo {author} {\bibfnamefont {R.-P.}\ \bibnamefont
  {Riwar}}, \bibinfo {author} {\bibfnamefont {A.}~\bibnamefont {Hosseinkhani}},
  \bibinfo {author} {\bibfnamefont {L.~D.}\ \bibnamefont {Burkhart}}, \bibinfo
  {author} {\bibfnamefont {Y.~Y.}\ \bibnamefont {Gao}}, \bibinfo {author}
  {\bibfnamefont {R.~J.}\ \bibnamefont {Schoelkopf}}, \bibinfo {author}
  {\bibfnamefont {L.~I.}\ \bibnamefont {Glazman}},\ and\ \bibinfo {author}
  {\bibfnamefont {G.}~\bibnamefont {Catelani}},\ }\bibfield  {title} {\bibinfo
  {title} {Normal-metal quasiparticle traps for superconducting qubits},\
  }\href {https://doi.org/10.1103/PhysRevB.94.104516} {\bibfield  {journal}
  {\bibinfo  {journal} {Phys. Rev. B}\ }\textbf {\bibinfo {volume} {94}},\
  \bibinfo {pages} {104516} (\bibinfo {year} {2016})}\BibitemShut {NoStop}%
\bibitem [{\citenamefont {Court}\ \emph {et~al.}(2008)\citenamefont {Court},
  \citenamefont {Ferguson}, \citenamefont {Lutchyn},\ and\ \citenamefont
  {Clark}}]{Court2008}%
  \BibitemOpen
  \bibfield  {author} {\bibinfo {author} {\bibfnamefont {N.~A.}\ \bibnamefont
  {Court}}, \bibinfo {author} {\bibfnamefont {A.~J.}\ \bibnamefont {Ferguson}},
  \bibinfo {author} {\bibfnamefont {R.}~\bibnamefont {Lutchyn}},\ and\ \bibinfo
  {author} {\bibfnamefont {R.~G.}\ \bibnamefont {Clark}},\ }\bibfield  {title}
  {\bibinfo {title} {Quantitative study of quasiparticle traps using the
  single-{{Cooper}}-pair transistor},\ }\href
  {https://doi.org/10.1103/PhysRevB.77.100501} {\bibfield  {journal} {\bibinfo
  {journal} {Phys. Rev. B}\ }\textbf {\bibinfo {volume} {77}},\ \bibinfo
  {pages} {100501(R)} (\bibinfo {year} {2008})}\BibitemShut {NoStop}%
\bibitem [{\citenamefont {Wang}\ \emph {et~al.}(2014)\citenamefont {Wang},
  \citenamefont {Gao}, \citenamefont {Pop}, \citenamefont {Vool}, \citenamefont
  {Axline}, \citenamefont {Brecht}, \citenamefont {Heeres}, \citenamefont
  {Frunzio}, \citenamefont {Devoret}, \citenamefont {Catelani}, \citenamefont
  {Glazman},\ and\ \citenamefont {Schoelkopf}}]{Wang2014}%
  \BibitemOpen
  \bibfield  {author} {\bibinfo {author} {\bibfnamefont {C.}~\bibnamefont
  {Wang}}, \bibinfo {author} {\bibfnamefont {Y.~Y.}\ \bibnamefont {Gao}},
  \bibinfo {author} {\bibfnamefont {I.~M.}\ \bibnamefont {Pop}}, \bibinfo
  {author} {\bibfnamefont {U.}~\bibnamefont {Vool}}, \bibinfo {author}
  {\bibfnamefont {C.}~\bibnamefont {Axline}}, \bibinfo {author} {\bibfnamefont
  {T.}~\bibnamefont {Brecht}}, \bibinfo {author} {\bibfnamefont {R.~W.}\
  \bibnamefont {Heeres}}, \bibinfo {author} {\bibfnamefont {L.}~\bibnamefont
  {Frunzio}}, \bibinfo {author} {\bibfnamefont {M.~H.}\ \bibnamefont
  {Devoret}}, \bibinfo {author} {\bibfnamefont {G.}~\bibnamefont {Catelani}},
  \bibinfo {author} {\bibfnamefont {L.~I.}\ \bibnamefont {Glazman}},\ and\
  \bibinfo {author} {\bibfnamefont {R.~J.}\ \bibnamefont {Schoelkopf}},\
  }\bibfield  {title} {\bibinfo {title} {Measurement and control of
  quasiparticle dynamics in a superconducting qubit},\ }\href
  {https://doi.org/10.1038/ncomms6836} {\bibfield  {journal} {\bibinfo
  {journal} {Nat Commun}\ }\textbf {\bibinfo {volume} {5}},\ \bibinfo {pages}
  {5836} (\bibinfo {year} {2014})}\BibitemShut {NoStop}%
\bibitem [{\citenamefont {Tinkham}(2004)}]{Tinkham2004}%
  \BibitemOpen
  \bibfield  {author} {\bibinfo {author} {\bibfnamefont {M.}~\bibnamefont
  {Tinkham}},\ }\href@noop {} {\emph {\bibinfo {title} {Introduction to
  {{Superconductivity}}}}}\ (\bibinfo  {publisher} {{Dover Publications}},\
  \bibinfo {year} {2004})\BibitemShut {NoStop}%
\bibitem [{\citenamefont {Rothwarf}\ and\ \citenamefont
  {Taylor}(1967)}]{Rothwarf1967}%
  \BibitemOpen
  \bibfield  {author} {\bibinfo {author} {\bibfnamefont {A.}~\bibnamefont
  {Rothwarf}}\ and\ \bibinfo {author} {\bibfnamefont {B.~N.}\ \bibnamefont
  {Taylor}},\ }\bibfield  {title} {\bibinfo {title} {Measurement of
  {{Recombination Lifetimes}} in {{Superconductors}}},\ }\href
  {https://doi.org/10.1103/PhysRevLett.19.27} {\bibfield  {journal} {\bibinfo
  {journal} {Phys. Rev. Lett.}\ }\textbf {\bibinfo {volume} {19}},\ \bibinfo
  {pages} {27} (\bibinfo {year} {1967})}\BibitemShut {NoStop}%
\bibitem [{\citenamefont {Kaplan}(1979)}]{Kaplan1979}%
  \BibitemOpen
  \bibfield  {author} {\bibinfo {author} {\bibfnamefont {S.~B.}\ \bibnamefont
  {Kaplan}},\ }\bibfield  {title} {\bibinfo {title} {Acoustic matching of
  superconducting films to substrates},\ }\href
  {https://doi.org/10.1007/BF00119193} {\bibfield  {journal} {\bibinfo
  {journal} {J Low Temp Phys}\ }\textbf {\bibinfo {volume} {37}},\ \bibinfo
  {pages} {343} (\bibinfo {year} {1979})}\BibitemShut {NoStop}%
\bibitem [{\citenamefont {Rostem}\ \emph {et~al.}(2018)\citenamefont {Rostem},
  \citenamefont {{de Visser}},\ and\ \citenamefont {Wollack}}]{Rostem2018}%
  \BibitemOpen
  \bibfield  {author} {\bibinfo {author} {\bibfnamefont {K.}~\bibnamefont
  {Rostem}}, \bibinfo {author} {\bibfnamefont {P.~J.}\ \bibnamefont {{de
  Visser}}},\ and\ \bibinfo {author} {\bibfnamefont {E.~J.}\ \bibnamefont
  {Wollack}},\ }\bibfield  {title} {\bibinfo {title} {Enhanced quasiparticle
  lifetime in a superconductor by selective blocking of recombination phonons
  with a phononic crystal},\ }\href
  {https://doi.org/10.1103/PhysRevB.98.014522} {\bibfield  {journal} {\bibinfo
  {journal} {Phys. Rev. B}\ }\textbf {\bibinfo {volume} {98}},\ \bibinfo
  {pages} {014522} (\bibinfo {year} {2018})}\BibitemShut {NoStop}%
\bibitem [{\citenamefont {Puurtinen}\ \emph {et~al.}(2020)\citenamefont
  {Puurtinen}, \citenamefont {Rostem}, \citenamefont {{de Visser}},\ and\
  \citenamefont {Maasilta}}]{Puurtinen2020}%
  \BibitemOpen
  \bibfield  {author} {\bibinfo {author} {\bibfnamefont {T.~A.}\ \bibnamefont
  {Puurtinen}}, \bibinfo {author} {\bibfnamefont {K.}~\bibnamefont {Rostem}},
  \bibinfo {author} {\bibfnamefont {P.~J.}\ \bibnamefont {{de Visser}}},\ and\
  \bibinfo {author} {\bibfnamefont {I.~J.}\ \bibnamefont {Maasilta}},\
  }\bibfield  {title} {\bibinfo {title} {A {{Composite Phononic Crystal
  Design}} for {{Quasiparticle Lifetime Enhancement}} in {{Kinetic Inductance
  Detectors}}},\ }\href {https://doi.org/10.1007/s10909-020-02423-4} {\bibfield
   {journal} {\bibinfo  {journal} {J Low Temp Phys}\ }\textbf {\bibinfo
  {volume} {199}},\ \bibinfo {pages} {577} (\bibinfo {year}
  {2020})}\BibitemShut {NoStop}%
\bibitem [{\citenamefont {Yates}\ \emph {et~al.}(2011)\citenamefont {Yates},
  \citenamefont {Baselmans}, \citenamefont {Endo}, \citenamefont {Janssen},
  \citenamefont {Ferrari}, \citenamefont {Diener},\ and\ \citenamefont
  {Baryshev}}]{Yates2011}%
  \BibitemOpen
  \bibfield  {author} {\bibinfo {author} {\bibfnamefont {S.~J.~C.}\
  \bibnamefont {Yates}}, \bibinfo {author} {\bibfnamefont {J.~J.~A.}\
  \bibnamefont {Baselmans}}, \bibinfo {author} {\bibfnamefont {A.}~\bibnamefont
  {Endo}}, \bibinfo {author} {\bibfnamefont {R.~M.~J.}\ \bibnamefont
  {Janssen}}, \bibinfo {author} {\bibfnamefont {L.}~\bibnamefont {Ferrari}},
  \bibinfo {author} {\bibfnamefont {P.}~\bibnamefont {Diener}},\ and\ \bibinfo
  {author} {\bibfnamefont {A.~M.}\ \bibnamefont {Baryshev}},\ }\bibfield
  {title} {\bibinfo {title} {Photon noise limited radiation detection with
  lens-antenna coupled microwave kinetic inductance detectors},\ }\href
  {https://doi.org/10.1063/1.3624846} {\bibfield  {journal} {\bibinfo
  {journal} {Appl. Phys. Lett.}\ }\textbf {\bibinfo {volume} {99}},\ \bibinfo
  {pages} {073505} (\bibinfo {year} {2011})}\BibitemShut {NoStop}%
\bibitem [{Note2()}]{Note2}%
  \BibitemOpen
  \bibinfo {note} {J. J .A. Baselmans et al., in preparation.}\BibitemShut
  {Stop}%
\bibitem [{\citenamefont {Noroozian}\ \emph {et~al.}(2009)\citenamefont
  {Noroozian}, \citenamefont {Gao}, \citenamefont {Zmuidzinas}, \citenamefont
  {LeDuc},\ and\ \citenamefont {Mazin}}]{Noroozian2009a}%
  \BibitemOpen
  \bibfield  {author} {\bibinfo {author} {\bibfnamefont {O.}~\bibnamefont
  {Noroozian}}, \bibinfo {author} {\bibfnamefont {J.}~\bibnamefont {Gao}},
  \bibinfo {author} {\bibfnamefont {J.}~\bibnamefont {Zmuidzinas}}, \bibinfo
  {author} {\bibfnamefont {H.~G.}\ \bibnamefont {LeDuc}},\ and\ \bibinfo
  {author} {\bibfnamefont {B.~A.}\ \bibnamefont {Mazin}},\ }\bibfield  {title}
  {\bibinfo {title} {Two-level system noise reduction for {{Microwave Kinetic
  Inductance Detectors}}},\ }\href {https://doi.org/10.1063/1.3292302}
  {\bibfield  {journal} {\bibinfo  {journal} {AIP Conference Proceedings}\
  }\textbf {\bibinfo {volume} {1185}},\ \bibinfo {pages} {148} (\bibinfo {year}
  {2009})}\BibitemShut {NoStop}%
\bibitem [{Note1()}]{Note1}%
  \BibitemOpen
  \bibinfo {note} {See Supplemental Material for all power spectral densities,
  device fabrication details, pulse rejection, responsivity calculations and
  influence of Al geometry and read power.}\BibitemShut {Stop}%
\bibitem [{\citenamefont {Baselmans}\ \emph {et~al.}(2012)\citenamefont
  {Baselmans}, \citenamefont {Yates}, \citenamefont {Diener},\ and\
  \citenamefont {{de Visser}}}]{Baselmans2012a}%
  \BibitemOpen
  \bibfield  {author} {\bibinfo {author} {\bibfnamefont {J.}~\bibnamefont
  {Baselmans}}, \bibinfo {author} {\bibfnamefont {S.}~\bibnamefont {Yates}},
  \bibinfo {author} {\bibfnamefont {P.}~\bibnamefont {Diener}},\ and\ \bibinfo
  {author} {\bibfnamefont {P.}~\bibnamefont {{de Visser}}},\ }\bibfield
  {title} {\bibinfo {title} {Ultra {{Low Background Cryogenic Test Facility}}
  for {{Far}}-{{Infrared Radiation Detectors}}},\ }\href
  {https://doi.org/10.1007/s10909-012-0511-0} {\bibfield  {journal} {\bibinfo
  {journal} {J Low Temp Phys}\ }\textbf {\bibinfo {volume} {167}},\ \bibinfo
  {pages} {360} (\bibinfo {year} {2012})}\BibitemShut {NoStop}%
\bibitem [{\citenamefont {Gao}\ \emph {et~al.}(2008{\natexlab{a}})\citenamefont
  {Gao}, \citenamefont {Zmuidzinas}, \citenamefont {Vayonakis}, \citenamefont
  {Day}, \citenamefont {Mazin},\ and\ \citenamefont {Leduc}}]{Gao2008}%
  \BibitemOpen
  \bibfield  {author} {\bibinfo {author} {\bibfnamefont {J.}~\bibnamefont
  {Gao}}, \bibinfo {author} {\bibfnamefont {J.}~\bibnamefont {Zmuidzinas}},
  \bibinfo {author} {\bibfnamefont {A.}~\bibnamefont {Vayonakis}}, \bibinfo
  {author} {\bibfnamefont {P.}~\bibnamefont {Day}}, \bibinfo {author}
  {\bibfnamefont {B.}~\bibnamefont {Mazin}},\ and\ \bibinfo {author}
  {\bibfnamefont {H.}~\bibnamefont {Leduc}},\ }\bibfield  {title} {\bibinfo
  {title} {Equivalence of the {{Effects}} on the {{Complex Conductivity}} of
  {{Superconductor}} due to {{Temperature Change}} and~{{External Pair
  Breaking}}},\ }\href {https://doi.org/10.1007/s10909-007-9688-z} {\bibfield
  {journal} {\bibinfo  {journal} {J Low Temp Phys}\ }\textbf {\bibinfo {volume}
  {151}},\ \bibinfo {pages} {557} (\bibinfo {year}
  {2008}{\natexlab{a}})}\BibitemShut {NoStop}%
\bibitem [{\citenamefont {{de Visser}}\ \emph {et~al.}(2021)\citenamefont {{de
  Visser}}, \citenamefont {{de Rooij}}, \citenamefont {Murugesan},
  \citenamefont {Thoen},\ and\ \citenamefont {Baselmans}}]{deVisser2021a}%
  \BibitemOpen
  \bibfield  {author} {\bibinfo {author} {\bibfnamefont {P.~J.}\ \bibnamefont
  {{de Visser}}}, \bibinfo {author} {\bibfnamefont {S.~A.}\ \bibnamefont {{de
  Rooij}}}, \bibinfo {author} {\bibfnamefont {V.}~\bibnamefont {Murugesan}},
  \bibinfo {author} {\bibfnamefont {D.~J.}\ \bibnamefont {Thoen}},\ and\
  \bibinfo {author} {\bibfnamefont {J.~J.}\ \bibnamefont {Baselmans}},\
  }\bibfield  {title} {\bibinfo {title} {Phonon-{{Trapping}}-{{Enhanced Energy
  Resolution}} in {{Superconducting Single}}-{{Photon Detectors}}},\ }\href
  {https://doi.org/10.1103/PhysRevApplied.16.034051} {\bibfield  {journal}
  {\bibinfo  {journal} {Phys. Rev. Applied}\ }\textbf {\bibinfo {volume}
  {16}},\ \bibinfo {pages} {034051} (\bibinfo {year} {2021})},\ \Eprint
  {https://arxiv.org/abs/2103.06723} {arXiv:2103.06723} \BibitemShut {NoStop}%
\bibitem [{\citenamefont {Abrikosov}\ and\ \citenamefont
  {Gor'kov}(1959{\natexlab{a}})}]{Abrikosov1959}%
  \BibitemOpen
  \bibfield  {author} {\bibinfo {author} {\bibfnamefont {A.~A.}\ \bibnamefont
  {Abrikosov}}\ and\ \bibinfo {author} {\bibfnamefont {L.~P.}\ \bibnamefont
  {Gor'kov}},\ }\bibfield  {title} {\bibinfo {title} {Superconducting alloys at
  finite temperatures},\ }\href@noop {} {\bibfield  {journal} {\bibinfo
  {journal} {Sov. Phys. - JETP (Engl. Transl.); (United States)}\ }\textbf
  {\bibinfo {volume} {9}},\ \bibinfo {pages} {220} (\bibinfo {year}
  {1959}{\natexlab{a}})}\BibitemShut {NoStop}%
\bibitem [{\citenamefont {Abrikosov}\ and\ \citenamefont
  {Gor'kov}(1959{\natexlab{b}})}]{Abrikosov1959a}%
  \BibitemOpen
  \bibfield  {author} {\bibinfo {author} {\bibfnamefont {A.~A.}\ \bibnamefont
  {Abrikosov}}\ and\ \bibinfo {author} {\bibfnamefont {L.~P.}\ \bibnamefont
  {Gor'kov}},\ }\bibfield  {title} {\bibinfo {title} {Theory of superconducting
  alloys. {{I}}. {{The}} electrodynamics of alloys at absolute zero},\
  }\href@noop {} {\bibfield  {journal} {\bibinfo  {journal} {Sov. Phys. - JETP
  (Engl. Transl.); (United States)}\ }\textbf {\bibinfo {volume} {8}},\
  \bibinfo {pages} {1090} (\bibinfo {year} {1959}{\natexlab{b}})}\BibitemShut
  {NoStop}%
\bibitem [{\citenamefont {{M{\"u}ller-Hartmann}}\ and\ \citenamefont
  {Zittartz}(1971)}]{Muller-Hartmann1971}%
  \BibitemOpen
  \bibfield  {author} {\bibinfo {author} {\bibfnamefont {E.}~\bibnamefont
  {{M{\"u}ller-Hartmann}}}\ and\ \bibinfo {author} {\bibfnamefont
  {J.}~\bibnamefont {Zittartz}},\ }\bibfield  {title} {\bibinfo {title} {Kondo
  {{Effect}} in {{Superconductors}}},\ }\href
  {https://doi.org/10.1103/PhysRevLett.26.428} {\bibfield  {journal} {\bibinfo
  {journal} {Phys. Rev. Lett.}\ }\textbf {\bibinfo {volume} {26}},\ \bibinfo
  {pages} {428} (\bibinfo {year} {1971})}\BibitemShut {NoStop}%
\bibitem [{\citenamefont {Yu}(1965)}]{Yu1965}%
  \BibitemOpen
  \bibfield  {author} {\bibinfo {author} {\bibfnamefont {L.}~\bibnamefont
  {Yu}},\ }\bibfield  {title} {\bibinfo {title} {{Bound State in
  Superconductors with Paramagnetic Impurities}},\ }\href@noop {} {\bibfield
  {journal} {\bibinfo  {journal} {Chin. J. Phys. (Peking) (Engl. Transl.)}\
  }\textbf {\bibinfo {volume} {21}},\ \bibinfo {pages} {75} (\bibinfo {year}
  {1965})}\BibitemShut {NoStop}%
\bibitem [{\citenamefont {Shiba}(1968)}]{Shiba1968}%
  \BibitemOpen
  \bibfield  {author} {\bibinfo {author} {\bibfnamefont {H.}~\bibnamefont
  {Shiba}},\ }\bibfield  {title} {\bibinfo {title} {Classical {{Spins}} in
  {{Superconductors}}},\ }\href {https://doi.org/10.1143/PTP.40.435} {\bibfield
   {journal} {\bibinfo  {journal} {Prog Theor Phys}\ }\textbf {\bibinfo
  {volume} {40}},\ \bibinfo {pages} {435} (\bibinfo {year} {1968})}\BibitemShut
  {NoStop}%
\bibitem [{\citenamefont {Rusinov}(1969)}]{Rusinov1969}%
  \BibitemOpen
  \bibfield  {author} {\bibinfo {author} {\bibfnamefont {A.~I.}\ \bibnamefont
  {Rusinov}},\ }\bibfield  {title} {\bibinfo {title} {On the {{Theory}} of
  {{Gapless Superconductivity}} in {{Alloys Containing Paramagnetic
  Impurities}}},\ }\href@noop {} {\bibfield  {journal} {\bibinfo  {journal}
  {Soviet Journal of Experimental and Theoretical Physics}\ }\textbf {\bibinfo
  {volume} {29}},\ \bibinfo {pages} {1101} (\bibinfo {year}
  {1969})}\BibitemShut {NoStop}%
\bibitem [{\citenamefont {Fominov}\ \emph {et~al.}(2011)\citenamefont
  {Fominov}, \citenamefont {Houzet},\ and\ \citenamefont
  {Glazman}}]{Fominov2011}%
  \BibitemOpen
  \bibfield  {author} {\bibinfo {author} {\bibfnamefont {Y.~V.}\ \bibnamefont
  {Fominov}}, \bibinfo {author} {\bibfnamefont {M.}~\bibnamefont {Houzet}},\
  and\ \bibinfo {author} {\bibfnamefont {L.~I.}\ \bibnamefont {Glazman}},\
  }\bibfield  {title} {\bibinfo {title} {Surface impedance of superconductors
  with weak magnetic impurities},\ }\href
  {https://doi.org/10.1103/PhysRevB.84.224517} {\bibfield  {journal} {\bibinfo
  {journal} {Phys. Rev. B}\ }\textbf {\bibinfo {volume} {84}},\ \bibinfo
  {pages} {224517} (\bibinfo {year} {2011})}\BibitemShut {NoStop}%
\bibitem [{\citenamefont {Kaiser}(1970)}]{Kaiser1970}%
  \BibitemOpen
  \bibfield  {author} {\bibinfo {author} {\bibfnamefont {A.~B.}\ \bibnamefont
  {Kaiser}},\ }\bibfield  {title} {\bibinfo {title} {Effect of non-magnetic
  localized states in superconducting alloys},\ }\href
  {https://doi.org/10.1088/0022-3719/3/2/023} {\bibfield  {journal} {\bibinfo
  {journal} {J. Phys. C: Solid State Phys.}\ }\textbf {\bibinfo {volume} {3}},\
  \bibinfo {pages} {410} (\bibinfo {year} {1970})}\BibitemShut {NoStop}%
\bibitem [{\citenamefont {Ghosal}\ \emph {et~al.}(2001)\citenamefont {Ghosal},
  \citenamefont {Randeria},\ and\ \citenamefont {Trivedi}}]{Ghosal2001}%
  \BibitemOpen
  \bibfield  {author} {\bibinfo {author} {\bibfnamefont {A.}~\bibnamefont
  {Ghosal}}, \bibinfo {author} {\bibfnamefont {M.}~\bibnamefont {Randeria}},\
  and\ \bibinfo {author} {\bibfnamefont {N.}~\bibnamefont {Trivedi}},\
  }\bibfield  {title} {\bibinfo {title} {Inhomogeneous pairing in highly
  disordered s-wave superconductors},\ }\href
  {https://doi.org/10.1103/PhysRevB.65.014501} {\bibfield  {journal} {\bibinfo
  {journal} {Phys. Rev. B}\ }\textbf {\bibinfo {volume} {65}},\ \bibinfo
  {pages} {014501} (\bibinfo {year} {2001})}\BibitemShut {NoStop}%
\bibitem [{\citenamefont {Bespalov}\ \emph
  {et~al.}(2016{\natexlab{a}})\citenamefont {Bespalov}, \citenamefont {Houzet},
  \citenamefont {Meyer},\ and\ \citenamefont {Nazarov}}]{Bespalov2016}%
  \BibitemOpen
  \bibfield  {author} {\bibinfo {author} {\bibfnamefont {A.}~\bibnamefont
  {Bespalov}}, \bibinfo {author} {\bibfnamefont {M.}~\bibnamefont {Houzet}},
  \bibinfo {author} {\bibfnamefont {J.~S.}\ \bibnamefont {Meyer}},\ and\
  \bibinfo {author} {\bibfnamefont {Y.~V.}\ \bibnamefont {Nazarov}},\
  }\bibfield  {title} {\bibinfo {title} {Density of states in gapped
  superconductors with pairing-potential impurities},\ }\href
  {https://doi.org/10.1103/PhysRevB.93.104521} {\bibfield  {journal} {\bibinfo
  {journal} {Phys. Rev. B}\ }\textbf {\bibinfo {volume} {93}},\ \bibinfo
  {pages} {104521} (\bibinfo {year} {2016}{\natexlab{a}})}\BibitemShut
  {NoStop}%
\bibitem [{\citenamefont {Bespalov}(2019)}]{Bespalov2019}%
  \BibitemOpen
  \bibfield  {author} {\bibinfo {author} {\bibfnamefont {A.~A.}\ \bibnamefont
  {Bespalov}},\ }\bibfield  {title} {\bibinfo {title} {Impurity-induced subgap
  states in superconductors with inhomogeneous pairing},\ }\href
  {https://doi.org/10.1103/PhysRevB.100.094507} {\bibfield  {journal} {\bibinfo
   {journal} {Phys. Rev. B}\ }\textbf {\bibinfo {volume} {100}},\ \bibinfo
  {pages} {094507} (\bibinfo {year} {2019})}\BibitemShut {NoStop}%
\bibitem [{\citenamefont {Larkin}\ and\ \citenamefont
  {Ovchinnikov}(1972)}]{Larkin1972}%
  \BibitemOpen
  \bibfield  {author} {\bibinfo {author} {\bibfnamefont {A.~I.}\ \bibnamefont
  {Larkin}}\ and\ \bibinfo {author} {\bibfnamefont {Y.~N.}\ \bibnamefont
  {Ovchinnikov}},\ }\bibfield  {title} {\bibinfo {title} {Density of {{States}}
  in {{Inhomogeneous Superconductors}}},\ }\href@noop {} {\bibfield  {journal}
  {\bibinfo  {journal} {Soviet Journal of Experimental and Theoretical
  Physics}\ }\textbf {\bibinfo {volume} {34}},\ \bibinfo {pages} {1144}
  (\bibinfo {year} {1972})}\BibitemShut {NoStop}%
\bibitem [{\citenamefont {Feigel'man}\ and\ \citenamefont
  {Skvortsov}(2012)}]{Feigelman2012}%
  \BibitemOpen
  \bibfield  {author} {\bibinfo {author} {\bibfnamefont {M.~V.}\ \bibnamefont
  {Feigel'man}}\ and\ \bibinfo {author} {\bibfnamefont {M.~A.}\ \bibnamefont
  {Skvortsov}},\ }\bibfield  {title} {\bibinfo {title} {Universal
  {{Broadening}} of the {{Bardeen}}-{{Cooper}}-{{Schrieffer Coherence Peak}} of
  {{Disordered Superconducting Films}}},\ }\href
  {https://doi.org/10.1103/PhysRevLett.109.147002} {\bibfield  {journal}
  {\bibinfo  {journal} {Phys. Rev. Lett.}\ }\textbf {\bibinfo {volume} {109}},\
  \bibinfo {pages} {147002} (\bibinfo {year} {2012})}\BibitemShut {NoStop}%
\bibitem [{\citenamefont {Chubov}\ \emph {et~al.}(1969)\citenamefont {Chubov},
  \citenamefont {Eremenko},\ and\ \citenamefont {Pilipenko}}]{Chubov1969}%
  \BibitemOpen
  \bibfield  {author} {\bibinfo {author} {\bibfnamefont {P.}~\bibnamefont
  {Chubov}}, \bibinfo {author} {\bibfnamefont {V.}~\bibnamefont {Eremenko}},\
  and\ \bibinfo {author} {\bibfnamefont {Y.~A.}\ \bibnamefont {Pilipenko}},\
  }\bibfield  {title} {\bibinfo {title} {Dependence of the critical temperature
  and energy gap on the thickness of superconducting aluminum films},\
  }\href@noop {} {\bibfield  {journal} {\bibinfo  {journal} {SOV PHYS JETP}\
  }\textbf {\bibinfo {volume} {28}},\ \bibinfo {pages} {389} (\bibinfo {year}
  {1969})}\BibitemShut {NoStop}%
\bibitem [{\citenamefont {Faoro}\ and\ \citenamefont
  {Ioffe}(2008)}]{Faoro2008}%
  \BibitemOpen
  \bibfield  {author} {\bibinfo {author} {\bibfnamefont {L.}~\bibnamefont
  {Faoro}}\ and\ \bibinfo {author} {\bibfnamefont {L.~B.}\ \bibnamefont
  {Ioffe}},\ }\bibfield  {title} {\bibinfo {title} {Microscopic {{Origin}} of
  {{Low}}-{{Frequency Flux Noise}} in {{Josephson Circuits}}},\ }\href
  {https://doi.org/10.1103/PhysRevLett.100.227005} {\bibfield  {journal}
  {\bibinfo  {journal} {Phys. Rev. Lett.}\ }\textbf {\bibinfo {volume} {100}},\
  \bibinfo {pages} {227005} (\bibinfo {year} {2008})}\BibitemShut {NoStop}%
\bibitem [{\citenamefont {Kumar}\ \emph {et~al.}(2016)\citenamefont {Kumar},
  \citenamefont {Sendelbach}, \citenamefont {Beck}, \citenamefont {Freeland},
  \citenamefont {Wang}, \citenamefont {Wang}, \citenamefont {Yu}, \citenamefont
  {Wu}, \citenamefont {Pappas},\ and\ \citenamefont {McDermott}}]{Kumar2016}%
  \BibitemOpen
  \bibfield  {author} {\bibinfo {author} {\bibfnamefont {P.}~\bibnamefont
  {Kumar}}, \bibinfo {author} {\bibfnamefont {S.}~\bibnamefont {Sendelbach}},
  \bibinfo {author} {\bibfnamefont {M.~A.}\ \bibnamefont {Beck}}, \bibinfo
  {author} {\bibfnamefont {J.~W.}\ \bibnamefont {Freeland}}, \bibinfo {author}
  {\bibfnamefont {Z.}~\bibnamefont {Wang}}, \bibinfo {author} {\bibfnamefont
  {H.}~\bibnamefont {Wang}}, \bibinfo {author} {\bibfnamefont {C.~C.}\
  \bibnamefont {Yu}}, \bibinfo {author} {\bibfnamefont {R.~Q.}\ \bibnamefont
  {Wu}}, \bibinfo {author} {\bibfnamefont {D.~P.}\ \bibnamefont {Pappas}},\
  and\ \bibinfo {author} {\bibfnamefont {R.}~\bibnamefont {McDermott}},\
  }\bibfield  {title} {\bibinfo {title} {Origin and {{Reduction}} of 1/f
  {{Magnetic Flux Noise}} in {{Superconducting Devices}}},\ }\href
  {https://doi.org/10.1103/PhysRevApplied.6.041001} {\bibfield  {journal}
  {\bibinfo  {journal} {Phys. Rev. Applied}\ }\textbf {\bibinfo {volume} {6}},\
  \bibinfo {pages} {041001(R)} (\bibinfo {year} {2016})}\BibitemShut {NoStop}%
\bibitem [{\citenamefont {Bespalov}\ \emph
  {et~al.}(2016{\natexlab{b}})\citenamefont {Bespalov}, \citenamefont {Houzet},
  \citenamefont {Meyer},\ and\ \citenamefont {Nazarov}}]{Bespalov2016a}%
  \BibitemOpen
  \bibfield  {author} {\bibinfo {author} {\bibfnamefont {A.}~\bibnamefont
  {Bespalov}}, \bibinfo {author} {\bibfnamefont {M.}~\bibnamefont {Houzet}},
  \bibinfo {author} {\bibfnamefont {J.~S.}\ \bibnamefont {Meyer}},\ and\
  \bibinfo {author} {\bibfnamefont {Y.~V.}\ \bibnamefont {Nazarov}},\
  }\bibfield  {title} {\bibinfo {title} {Theoretical {{Model}} to {{Explain
  Excess}} of {{Quasiparticles}} in {{Superconductors}}},\ }\href
  {https://doi.org/10.1103/PhysRevLett.117.117002} {\bibfield  {journal}
  {\bibinfo  {journal} {Phys. Rev. Lett.}\ }\textbf {\bibinfo {volume} {117}},\
  \bibinfo {pages} {117002} (\bibinfo {year} {2016}{\natexlab{b}})}\BibitemShut
  {NoStop}%
\bibitem [{\citenamefont {Meyer}\ \emph {et~al.}(2020)\citenamefont {Meyer},
  \citenamefont {Houzet},\ and\ \citenamefont {Nazarov}}]{Meyer2020}%
  \BibitemOpen
  \bibfield  {author} {\bibinfo {author} {\bibfnamefont {J.~S.}\ \bibnamefont
  {Meyer}}, \bibinfo {author} {\bibfnamefont {M.}~\bibnamefont {Houzet}},\ and\
  \bibinfo {author} {\bibfnamefont {Y.~V.}\ \bibnamefont {Nazarov}},\
  }\bibfield  {title} {\bibinfo {title} {Dynamical {{Spin Polarization}} of
  {{Excess Quasiparticles}} in {{Superconductors}}},\ }\href
  {https://doi.org/10.1103/PhysRevLett.125.097006} {\bibfield  {journal}
  {\bibinfo  {journal} {Phys. Rev. Lett.}\ }\textbf {\bibinfo {volume} {125}},\
  \bibinfo {pages} {097006} (\bibinfo {year} {2020})}\BibitemShut {NoStop}%
\bibitem [{\citenamefont {Kozorezov}\ \emph
  {et~al.}(2008{\natexlab{a}})\citenamefont {Kozorezov}, \citenamefont
  {Golubov}, \citenamefont {Wigmore}, \citenamefont {Martin}, \citenamefont
  {Verhoeve}, \citenamefont {Hijmering},\ and\ \citenamefont
  {Jerjen}}]{Kozorezov2008}%
  \BibitemOpen
  \bibfield  {author} {\bibinfo {author} {\bibfnamefont {A.~G.}\ \bibnamefont
  {Kozorezov}}, \bibinfo {author} {\bibfnamefont {A.~A.}\ \bibnamefont
  {Golubov}}, \bibinfo {author} {\bibfnamefont {J.~K.}\ \bibnamefont
  {Wigmore}}, \bibinfo {author} {\bibfnamefont {D.}~\bibnamefont {Martin}},
  \bibinfo {author} {\bibfnamefont {P.}~\bibnamefont {Verhoeve}}, \bibinfo
  {author} {\bibfnamefont {R.~A.}\ \bibnamefont {Hijmering}},\ and\ \bibinfo
  {author} {\bibfnamefont {I.}~\bibnamefont {Jerjen}},\ }\bibfield  {title}
  {\bibinfo {title} {Inelastic scattering of quasiparticles in a superconductor
  with magnetic impurities},\ }\href
  {https://doi.org/10.1103/PhysRevB.78.174501} {\bibfield  {journal} {\bibinfo
  {journal} {Phys. Rev. B}\ }\textbf {\bibinfo {volume} {78}},\ \bibinfo
  {pages} {174501} (\bibinfo {year} {2008}{\natexlab{a}})}\BibitemShut
  {NoStop}%
\bibitem [{\citenamefont {Kozorezov}\ \emph {et~al.}(2009)\citenamefont
  {Kozorezov}, \citenamefont {Golubov}, \citenamefont {Wigmore}, \citenamefont
  {Martin}, \citenamefont {Verhoeve}, \citenamefont {Hijmering},\ and\
  \citenamefont {Jerjen}}]{Kozorezov2009}%
  \BibitemOpen
  \bibfield  {author} {\bibinfo {author} {\bibfnamefont {A.~G.}\ \bibnamefont
  {Kozorezov}}, \bibinfo {author} {\bibfnamefont {A.~A.}\ \bibnamefont
  {Golubov}}, \bibinfo {author} {\bibfnamefont {J.~K.}\ \bibnamefont
  {Wigmore}}, \bibinfo {author} {\bibfnamefont {D.}~\bibnamefont {Martin}},
  \bibinfo {author} {\bibfnamefont {P.}~\bibnamefont {Verhoeve}}, \bibinfo
  {author} {\bibfnamefont {R.~A.}\ \bibnamefont {Hijmering}},\ and\ \bibinfo
  {author} {\bibfnamefont {I.}~\bibnamefont {Jerjen}},\ }\bibfield  {title}
  {\bibinfo {title} {The {{Effect}} of {{Magnetic Impurities}} on the
  {{Response}} of {{Superconducting Photon Detectors}}},\ }\href
  {https://doi.org/10.1109/TASC.2009.2018062} {\bibfield  {journal} {\bibinfo
  {journal} {IEEE Transactions on Applied Superconductivity}\ }\textbf
  {\bibinfo {volume} {19}},\ \bibinfo {pages} {440} (\bibinfo {year}
  {2009})}\BibitemShut {NoStop}%
\bibitem [{\citenamefont {Kozorezov}\ \emph {et~al.}(2001)\citenamefont
  {Kozorezov}, \citenamefont {Wigmore}, \citenamefont {Peacock}, \citenamefont
  {Poelaert}, \citenamefont {Verhoeve}, \citenamefont {{den Hartog}},\ and\
  \citenamefont {Brammertz}}]{Kozorezov2001}%
  \BibitemOpen
  \bibfield  {author} {\bibinfo {author} {\bibfnamefont {A.~G.}\ \bibnamefont
  {Kozorezov}}, \bibinfo {author} {\bibfnamefont {J.~K.}\ \bibnamefont
  {Wigmore}}, \bibinfo {author} {\bibfnamefont {A.}~\bibnamefont {Peacock}},
  \bibinfo {author} {\bibfnamefont {A.}~\bibnamefont {Poelaert}}, \bibinfo
  {author} {\bibfnamefont {P.}~\bibnamefont {Verhoeve}}, \bibinfo {author}
  {\bibfnamefont {R.}~\bibnamefont {{den Hartog}}},\ and\ \bibinfo {author}
  {\bibfnamefont {G.}~\bibnamefont {Brammertz}},\ }\bibfield  {title} {\bibinfo
  {title} {Local trap spectroscopy in superconducting tunnel junctions},\
  }\href {https://doi.org/10.1063/1.1377624} {\bibfield  {journal} {\bibinfo
  {journal} {Appl. Phys. Lett.}\ }\textbf {\bibinfo {volume} {78}},\ \bibinfo
  {pages} {3654} (\bibinfo {year} {2001})}\BibitemShut {NoStop}%
\bibitem [{\citenamefont {Kozorezov}\ \emph
  {et~al.}(2008{\natexlab{b}})\citenamefont {Kozorezov}, \citenamefont
  {Hijmering}, \citenamefont {Brammertz}, \citenamefont {Wigmore},
  \citenamefont {Peacock}, \citenamefont {Martin}, \citenamefont {Verhoeve},
  \citenamefont {Golubov},\ and\ \citenamefont {Rogalla}}]{Kozorezov2008a}%
  \BibitemOpen
  \bibfield  {author} {\bibinfo {author} {\bibfnamefont {A.~G.}\ \bibnamefont
  {Kozorezov}}, \bibinfo {author} {\bibfnamefont {R.~A.}\ \bibnamefont
  {Hijmering}}, \bibinfo {author} {\bibfnamefont {G.}~\bibnamefont
  {Brammertz}}, \bibinfo {author} {\bibfnamefont {J.~K.}\ \bibnamefont
  {Wigmore}}, \bibinfo {author} {\bibfnamefont {A.}~\bibnamefont {Peacock}},
  \bibinfo {author} {\bibfnamefont {D.}~\bibnamefont {Martin}}, \bibinfo
  {author} {\bibfnamefont {P.}~\bibnamefont {Verhoeve}}, \bibinfo {author}
  {\bibfnamefont {A.~A.}\ \bibnamefont {Golubov}},\ and\ \bibinfo {author}
  {\bibfnamefont {H.}~\bibnamefont {Rogalla}},\ }\bibfield  {title} {\bibinfo
  {title} {Dynamics of nonequilibrium quasiparticles in narrow-gap
  superconducting tunnel junctions},\ }\href
  {https://doi.org/10.1103/PhysRevB.77.014501} {\bibfield  {journal} {\bibinfo
  {journal} {Phys. Rev. B}\ }\textbf {\bibinfo {volume} {77}},\ \bibinfo
  {pages} {014501} (\bibinfo {year} {2008}{\natexlab{b}})}\BibitemShut
  {NoStop}%
\bibitem [{\citenamefont {Bueno}\ \emph {et~al.}(2014)\citenamefont {Bueno},
  \citenamefont {Coumou}, \citenamefont {Zheng}, \citenamefont {{de Visser}},
  \citenamefont {Klapwijk}, \citenamefont {Driessen}, \citenamefont {Doyle},\
  and\ \citenamefont {Baselmans}}]{Bueno2014}%
  \BibitemOpen
  \bibfield  {author} {\bibinfo {author} {\bibfnamefont {J.}~\bibnamefont
  {Bueno}}, \bibinfo {author} {\bibfnamefont {P.~C. J.~J.}\ \bibnamefont
  {Coumou}}, \bibinfo {author} {\bibfnamefont {G.}~\bibnamefont {Zheng}},
  \bibinfo {author} {\bibfnamefont {P.~J.}\ \bibnamefont {{de Visser}}},
  \bibinfo {author} {\bibfnamefont {T.~M.}\ \bibnamefont {Klapwijk}}, \bibinfo
  {author} {\bibfnamefont {E.~F.~C.}\ \bibnamefont {Driessen}}, \bibinfo
  {author} {\bibfnamefont {S.}~\bibnamefont {Doyle}},\ and\ \bibinfo {author}
  {\bibfnamefont {J.~J.~A.}\ \bibnamefont {Baselmans}},\ }\bibfield  {title}
  {\bibinfo {title} {Anomalous response of superconducting titanium nitride
  resonators to terahertz radiation},\ }\href
  {https://doi.org/10.1063/1.4901536} {\bibfield  {journal} {\bibinfo
  {journal} {Appl. Phys. Lett.}\ }\textbf {\bibinfo {volume} {105}},\ \bibinfo
  {pages} {192601} (\bibinfo {year} {2014})}\BibitemShut {NoStop}%
\bibitem [{\citenamefont {Gr{\"u}nhaupt}\ \emph {et~al.}(2018)\citenamefont
  {Gr{\"u}nhaupt}, \citenamefont {Maleeva}, \citenamefont {Skacel},
  \citenamefont {Calvo}, \citenamefont {{Levy-Bertrand}}, \citenamefont
  {Ustinov}, \citenamefont {Rotzinger}, \citenamefont {Monfardini},
  \citenamefont {Catelani},\ and\ \citenamefont {Pop}}]{Grunhaupt2018}%
  \BibitemOpen
  \bibfield  {author} {\bibinfo {author} {\bibfnamefont {L.}~\bibnamefont
  {Gr{\"u}nhaupt}}, \bibinfo {author} {\bibfnamefont {N.}~\bibnamefont
  {Maleeva}}, \bibinfo {author} {\bibfnamefont {S.~T.}\ \bibnamefont {Skacel}},
  \bibinfo {author} {\bibfnamefont {M.}~\bibnamefont {Calvo}}, \bibinfo
  {author} {\bibfnamefont {F.}~\bibnamefont {{Levy-Bertrand}}}, \bibinfo
  {author} {\bibfnamefont {A.~V.}\ \bibnamefont {Ustinov}}, \bibinfo {author}
  {\bibfnamefont {H.}~\bibnamefont {Rotzinger}}, \bibinfo {author}
  {\bibfnamefont {A.}~\bibnamefont {Monfardini}}, \bibinfo {author}
  {\bibfnamefont {G.}~\bibnamefont {Catelani}},\ and\ \bibinfo {author}
  {\bibfnamefont {I.~M.}\ \bibnamefont {Pop}},\ }\bibfield  {title} {\bibinfo
  {title} {Loss {{Mechanisms}} and {{Quasiparticle Dynamics}} in
  {{Superconducting Microwave Resonators Made}} of {{Thin}}-{{Film Granular
  Aluminum}}},\ }\href {https://doi.org/10.1103/PhysRevLett.121.117001}
  {\bibfield  {journal} {\bibinfo  {journal} {Phys. Rev. Lett.}\ }\textbf
  {\bibinfo {volume} {121}},\ \bibinfo {pages} {117001} (\bibinfo {year}
  {2018})}\BibitemShut {NoStop}%
\bibitem [{\citenamefont {{de Rooij}}(2020)}]{deRooij2020}%
  \BibitemOpen
  \bibfield  {author} {\bibinfo {author} {\bibfnamefont {S.~A.~H.}\
  \bibnamefont {{de Rooij}}},\ }\emph {\bibinfo {title} {Quasiparticle
  {{Dynamics}} in {{Optical MKIDs}}: Single {{Photon Response}} and
  {{Temperture Dependent Generation}}-{{Recombination Noise}}}},\ \href@noop {}
  {Master's thesis},\ \bibinfo  {school} {Delft University of Technology},
  \bibinfo {address} {{Delft}} (\bibinfo {year} {2020})\BibitemShut {NoStop}%
\bibitem [{\citenamefont {Day}\ \emph {et~al.}(2003)\citenamefont {Day},
  \citenamefont {LeDuc}, \citenamefont {Mazin}, \citenamefont {Vayonakis},\
  and\ \citenamefont {Zmuidzinas}}]{Day2003}%
  \BibitemOpen
  \bibfield  {author} {\bibinfo {author} {\bibfnamefont {P.~K.}\ \bibnamefont
  {Day}}, \bibinfo {author} {\bibfnamefont {H.~G.}\ \bibnamefont {LeDuc}},
  \bibinfo {author} {\bibfnamefont {B.~A.}\ \bibnamefont {Mazin}}, \bibinfo
  {author} {\bibfnamefont {A.}~\bibnamefont {Vayonakis}},\ and\ \bibinfo
  {author} {\bibfnamefont {J.}~\bibnamefont {Zmuidzinas}},\ }\bibfield  {title}
  {\bibinfo {title} {A broadband superconducting detector suitable for use in
  large arrays},\ }\href {https://doi.org/10.1038/nature02037} {\bibfield
  {journal} {\bibinfo  {journal} {Nature}\ }\textbf {\bibinfo {volume} {425}},\
  \bibinfo {pages} {817} (\bibinfo {year} {2003})}\BibitemShut {NoStop}%
\bibitem [{\citenamefont {Zmuidzinas}(2012)}]{Zmuidzinas2012}%
  \BibitemOpen
  \bibfield  {author} {\bibinfo {author} {\bibfnamefont {J.}~\bibnamefont
  {Zmuidzinas}},\ }\bibfield  {title} {\bibinfo {title} {Superconducting
  {{Microresonators}}: Physics and {{Applications}}},\ }\href
  {https://doi.org/10.1146/annurev-conmatphys-020911-125022} {\bibfield
  {journal} {\bibinfo  {journal} {Annu. Rev. Condens. Matter Phys.}\ }\textbf
  {\bibinfo {volume} {3}},\ \bibinfo {pages} {169} (\bibinfo {year}
  {2012})}\BibitemShut {NoStop}%
\bibitem [{\citenamefont {Thoen}\ \emph {et~al.}(2017)\citenamefont {Thoen},
  \citenamefont {Bos}, \citenamefont {Haalebos}, \citenamefont {Klapwijk},
  \citenamefont {Baselmans},\ and\ \citenamefont {Endo}}]{Thoen2017}%
  \BibitemOpen
  \bibfield  {author} {\bibinfo {author} {\bibfnamefont {D.~J.}\ \bibnamefont
  {Thoen}}, \bibinfo {author} {\bibfnamefont {B.~G.~C.}\ \bibnamefont {Bos}},
  \bibinfo {author} {\bibfnamefont {E.~A.~F.}\ \bibnamefont {Haalebos}},
  \bibinfo {author} {\bibfnamefont {T.~M.}\ \bibnamefont {Klapwijk}}, \bibinfo
  {author} {\bibfnamefont {J.~J.~A.}\ \bibnamefont {Baselmans}},\ and\ \bibinfo
  {author} {\bibfnamefont {A.}~\bibnamefont {Endo}},\ }\bibfield  {title}
  {\bibinfo {title} {Superconducting {{NbTin Thin Films With Highly Uniform
  Properties Over}} a 100 mm {{Wafer}}},\ }\href
  {https://doi.org/10.1109/TASC.2016.2631948} {\bibfield  {journal} {\bibinfo
  {journal} {IEEE Transactions on Applied Superconductivity}\ }\textbf
  {\bibinfo {volume} {27}},\ \bibinfo {pages} {1} (\bibinfo {year}
  {2017})}\BibitemShut {NoStop}%
\bibitem [{\citenamefont {Gao}\ \emph {et~al.}(2008{\natexlab{b}})\citenamefont
  {Gao}, \citenamefont {Daal}, \citenamefont {Vayonakis}, \citenamefont
  {Kumar}, \citenamefont {Zmuidzinas}, \citenamefont {Sadoulet}, \citenamefont
  {Mazin}, \citenamefont {Day},\ and\ \citenamefont {Leduc}}]{Gao2008b}%
  \BibitemOpen
  \bibfield  {author} {\bibinfo {author} {\bibfnamefont {J.}~\bibnamefont
  {Gao}}, \bibinfo {author} {\bibfnamefont {M.}~\bibnamefont {Daal}}, \bibinfo
  {author} {\bibfnamefont {A.}~\bibnamefont {Vayonakis}}, \bibinfo {author}
  {\bibfnamefont {S.}~\bibnamefont {Kumar}}, \bibinfo {author} {\bibfnamefont
  {J.}~\bibnamefont {Zmuidzinas}}, \bibinfo {author} {\bibfnamefont
  {B.}~\bibnamefont {Sadoulet}}, \bibinfo {author} {\bibfnamefont {B.~A.}\
  \bibnamefont {Mazin}}, \bibinfo {author} {\bibfnamefont {P.~K.}\ \bibnamefont
  {Day}},\ and\ \bibinfo {author} {\bibfnamefont {H.~G.}\ \bibnamefont
  {Leduc}},\ }\bibfield  {title} {\bibinfo {title} {Experimental evidence for a
  surface distribution of two-level systems in superconducting lithographed
  microwave resonators},\ }\href {https://doi.org/10.1063/1.2906373} {\bibfield
   {journal} {\bibinfo  {journal} {Appl. Phys. Lett.}\ }\textbf {\bibinfo
  {volume} {92}},\ \bibinfo {pages} {152505} (\bibinfo {year}
  {2008}{\natexlab{b}})}\BibitemShut {NoStop}%
\bibitem [{\citenamefont {{de Visser}}\ \emph
  {et~al.}(2014{\natexlab{b}})\citenamefont {{de Visser}}, \citenamefont
  {Baselmans}, \citenamefont {Bueno}, \citenamefont {Llombart},\ and\
  \citenamefont {Klapwijk}}]{deVisser2014b}%
  \BibitemOpen
  \bibfield  {author} {\bibinfo {author} {\bibfnamefont {P.~J.}\ \bibnamefont
  {{de Visser}}}, \bibinfo {author} {\bibfnamefont {J.~J.~A.}\ \bibnamefont
  {Baselmans}}, \bibinfo {author} {\bibfnamefont {J.}~\bibnamefont {Bueno}},
  \bibinfo {author} {\bibfnamefont {N.}~\bibnamefont {Llombart}},\ and\
  \bibinfo {author} {\bibfnamefont {T.~M.}\ \bibnamefont {Klapwijk}},\
  }\bibfield  {title} {\bibinfo {title} {Fluctuations in the electron system of
  a superconductor exposed to a photon flux},\ }\href
  {https://doi.org/10.1038/ncomms4130} {\bibfield  {journal} {\bibinfo
  {journal} {Nature Communications}\ }\textbf {\bibinfo {volume} {5}},\
  \bibinfo {pages} {3130} (\bibinfo {year} {2014}{\natexlab{b}})}\BibitemShut
  {NoStop}%
\bibitem [{\citenamefont {Swenson}\ \emph {et~al.}(2013)\citenamefont
  {Swenson}, \citenamefont {Day}, \citenamefont {Eom}, \citenamefont {Leduc},
  \citenamefont {Llombart}, \citenamefont {McKenney}, \citenamefont
  {Noroozian},\ and\ \citenamefont {Zmuidzinas}}]{Swenson2013}%
  \BibitemOpen
  \bibfield  {author} {\bibinfo {author} {\bibfnamefont {L.~J.}\ \bibnamefont
  {Swenson}}, \bibinfo {author} {\bibfnamefont {P.~K.}\ \bibnamefont {Day}},
  \bibinfo {author} {\bibfnamefont {B.~H.}\ \bibnamefont {Eom}}, \bibinfo
  {author} {\bibfnamefont {H.~G.}\ \bibnamefont {Leduc}}, \bibinfo {author}
  {\bibfnamefont {N.}~\bibnamefont {Llombart}}, \bibinfo {author}
  {\bibfnamefont {C.~M.}\ \bibnamefont {McKenney}}, \bibinfo {author}
  {\bibfnamefont {O.}~\bibnamefont {Noroozian}},\ and\ \bibinfo {author}
  {\bibfnamefont {J.}~\bibnamefont {Zmuidzinas}},\ }\bibfield  {title}
  {\bibinfo {title} {Operation of a titanium nitride superconducting
  microresonator detector in the nonlinear regime},\ }\href
  {https://doi.org/10.1063/1.4794808} {\bibfield  {journal} {\bibinfo
  {journal} {Journal of Applied Physics}\ }\textbf {\bibinfo {volume} {113}},\
  \bibinfo {pages} {104501} (\bibinfo {year} {2013})}\BibitemShut {NoStop}%
\bibitem [{\citenamefont {Valenti}\ \emph {et~al.}(2019)\citenamefont
  {Valenti}, \citenamefont {Henriques}, \citenamefont {Catelani}, \citenamefont
  {Maleeva}, \citenamefont {Gr{\"u}nhaupt}, \citenamefont {{von L{\"u}pke}},
  \citenamefont {Skacel}, \citenamefont {Winkel}, \citenamefont {Bilmes},
  \citenamefont {Ustinov}, \citenamefont {Goupy}, \citenamefont {Calvo},
  \citenamefont {Beno{\^i}t}, \citenamefont {{Levy-Bertrand}}, \citenamefont
  {Monfardini},\ and\ \citenamefont {Pop}}]{Valenti2019}%
  \BibitemOpen
  \bibfield  {author} {\bibinfo {author} {\bibfnamefont {F.}~\bibnamefont
  {Valenti}}, \bibinfo {author} {\bibfnamefont {F.}~\bibnamefont {Henriques}},
  \bibinfo {author} {\bibfnamefont {G.}~\bibnamefont {Catelani}}, \bibinfo
  {author} {\bibfnamefont {N.}~\bibnamefont {Maleeva}}, \bibinfo {author}
  {\bibfnamefont {L.}~\bibnamefont {Gr{\"u}nhaupt}}, \bibinfo {author}
  {\bibfnamefont {U.}~\bibnamefont {{von L{\"u}pke}}}, \bibinfo {author}
  {\bibfnamefont {S.~T.}\ \bibnamefont {Skacel}}, \bibinfo {author}
  {\bibfnamefont {P.}~\bibnamefont {Winkel}}, \bibinfo {author} {\bibfnamefont
  {A.}~\bibnamefont {Bilmes}}, \bibinfo {author} {\bibfnamefont {A.~V.}\
  \bibnamefont {Ustinov}}, \bibinfo {author} {\bibfnamefont {J.}~\bibnamefont
  {Goupy}}, \bibinfo {author} {\bibfnamefont {M.}~\bibnamefont {Calvo}},
  \bibinfo {author} {\bibfnamefont {A.}~\bibnamefont {Beno{\^i}t}}, \bibinfo
  {author} {\bibfnamefont {F.}~\bibnamefont {{Levy-Bertrand}}}, \bibinfo
  {author} {\bibfnamefont {A.}~\bibnamefont {Monfardini}},\ and\ \bibinfo
  {author} {\bibfnamefont {I.~M.}\ \bibnamefont {Pop}},\ }\bibfield  {title}
  {\bibinfo {title} {Interplay {{Between Kinetic Inductance}},
  {{Nonlinearity}}, and {{Quasiparticle Dynamics}} in {{Granular Aluminum
  Microwave Kinetic Inductance Detectors}}},\ }\href
  {https://doi.org/10.1103/PhysRevApplied.11.054087} {\bibfield  {journal}
  {\bibinfo  {journal} {Phys. Rev. Applied}\ }\textbf {\bibinfo {volume}
  {11}},\ \bibinfo {pages} {054087} (\bibinfo {year} {2019})}\BibitemShut
  {NoStop}%
\bibitem [{\citenamefont {Thomas}\ \emph {et~al.}(2020)\citenamefont {Thomas},
  \citenamefont {Withington}, \citenamefont {Sun}, \citenamefont {Skyrme},\
  and\ \citenamefont {Goldie}}]{Thomas2020}%
  \BibitemOpen
  \bibfield  {author} {\bibinfo {author} {\bibfnamefont {C.~N.}\ \bibnamefont
  {Thomas}}, \bibinfo {author} {\bibfnamefont {S.}~\bibnamefont {Withington}},
  \bibinfo {author} {\bibfnamefont {Z.}~\bibnamefont {Sun}}, \bibinfo {author}
  {\bibfnamefont {T.}~\bibnamefont {Skyrme}},\ and\ \bibinfo {author}
  {\bibfnamefont {D.~J.}\ \bibnamefont {Goldie}},\ }\bibfield  {title}
  {\bibinfo {title} {Nonlinear effects in superconducting thin film microwave
  resonators},\ }\href {https://doi.org/10.1088/1367-2630/ab97e8} {\bibfield
  {journal} {\bibinfo  {journal} {New J. Phys.}\ }\textbf {\bibinfo {volume}
  {22}},\ \bibinfo {pages} {073028} (\bibinfo {year} {2020})},\ \Eprint
  {https://arxiv.org/abs/2001.02540} {arXiv:2001.02540} \BibitemShut {NoStop}%
\bibitem [{\citenamefont {{de Visser}}(2014)}]{deVisser2014}%
  \BibitemOpen
  \bibfield  {author} {\bibinfo {author} {\bibfnamefont {P.~J.}\ \bibnamefont
  {{de Visser}}},\ }\emph {\bibinfo {title} {Quasiparticle Dynamics in
  Aluminium Superconducting Microwave Resonators}},\ \href@noop {} {Ph.D.
  thesis},\ \bibinfo  {school} {Delft University of Technology}, \bibinfo
  {address} {{Delft}} (\bibinfo {year} {2014})\BibitemShut {NoStop}%
\bibitem [{\citenamefont {Guruswamy}\ \emph {et~al.}(2014)\citenamefont
  {Guruswamy}, \citenamefont {Goldie},\ and\ \citenamefont
  {Withington}}]{Guruswamy2014}%
  \BibitemOpen
  \bibfield  {author} {\bibinfo {author} {\bibfnamefont {T.}~\bibnamefont
  {Guruswamy}}, \bibinfo {author} {\bibfnamefont {D.~J.}\ \bibnamefont
  {Goldie}},\ and\ \bibinfo {author} {\bibfnamefont {S.}~\bibnamefont
  {Withington}},\ }\bibfield  {title} {\bibinfo {title} {Quasiparticle
  generation efficiency in superconducting thin films},\ }\href
  {https://doi.org/10.1088/0953-2048/27/5/055012} {\bibfield  {journal}
  {\bibinfo  {journal} {Supercond. Sci. Technol.}\ }\textbf {\bibinfo {volume}
  {27}},\ \bibinfo {pages} {055012} (\bibinfo {year} {2014})}\BibitemShut
  {NoStop}%
\end{thebibliography}%

\appendix
\onecolumngrid
\renewcommand\thefigure{S\arabic{figure}} 
\renewcommand\thetable{S\Roman{table}} 
\setcounter{figure}{0}
\setcounter{table}{0}
\section{SUPPLEMENTAL MATERIAL}
\section{Device Design and Fabrication}
The capacitive part is a (8-40-8)-$\mu m$ co-planar waveguide (CPW) with interdigitated capacitors (IDC) on both sides \cite{Noroozian2009a} \footnotemark[2], patterned in a \SI{200}{\nano\meter} thick NbTiN film \cite{Thoen2017} (normal state resistivity, $\rho_N=\SI{245.6}{\micro\ohm\centi\meter}$, $T_c=\SI{15.6}{\kelvin}$, residual resistance ratio, $RRR=0.93$). The large width of the IDC suppresses TLS noise, as the electric field is distributed over a larger area \cite{Gao2008b}. The capacitive fingers increase the device capacitance, allowing a resonator with a much lower total inductance for the same resonant frequency ($f_0\propto1/\sqrt{LC}$).\\
The sensitive part is a (23-1.73-23)-$\mu m$ CPW, with an Al ($\rho_N=\SI{1.6}{\micro\ohm\centi\meter}$, $T_c=\SI{1.26}{\kelvin}$, $RRR=4.0$) central line, which is \SI{312}{\micro\meter} long, \SI{50}{\nano\meter} thick and shorted to the NbTiN ground plane to make a quarter-wave resonator.\\
The NbTiN and Al films are sputtered on a \SI{350}{\micro\meter} thick Si wafer with a \SI{150}{\nano\meter} SiN film on top, deposited with low pressure chemical vapour deposition. A backside KOH etch is used to release the SiN membrane, which supports two of the four Al sections as seen in \cref{fig:LT165} of the main text. The final membrane thickness is \SI{110}{\nano\meter} due to over-etching. The SiN beneath the capacitive part is etched away to suppress two level system (TLS) noise.\\

\subsubsection{Quality Factors}
\begin{table}[!ht]
	\caption{The internal ($Q_i$), coupling ($Q_c$), loaded ($Q$) quality-factors and resonance frequency ($f_0$) at \SI{50}{\milli\kelvin}, measured from $S_{21}$ curves. An on-chip microwave power of \SI{-99}{\decibel m} was used to probe the resonators.}\label{tab:Qfactors}
	\begin{tabular}{ccc}
		\hline
		\hline
		& Membrane & Substrate \\
		\hline
		$Q_i$ & \SI{5.8e5}{} & \SI{1.0e6}{} \\
		$Q_c$ & \SI{4.6e4}{} & \SI{3.6e4}{} \\
		$Q$ & \SI{4.2e4}{} & \SI{3.5e4}{}\\
		\hline
		$f_0$ & \SI{4.3}{\giga\hertz} & \SI{4.5}{\giga\hertz}\\
		\hline
		\hline
	\end{tabular}
\end{table}
 \newpage
\section{Amplitude, Phase and Cross-Power Spectral Densities}
\begin{figure}[!ht]
	\includegraphics[width=.7\linewidth]{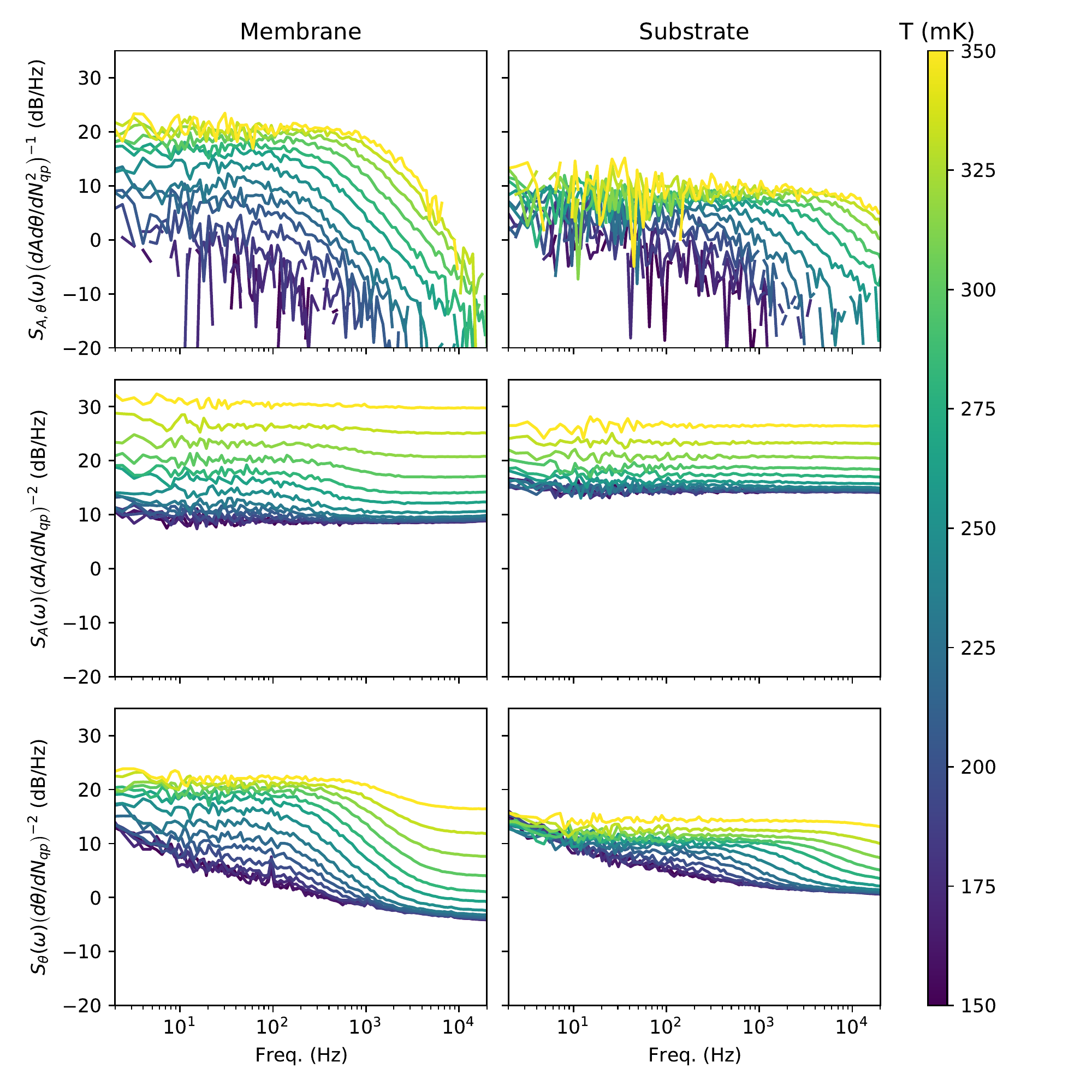}
	\caption{Cross-, amplitude and phase Power Spectral Densities (PSDs) for the membrane and substrate resonators (see titles) and different temperatures (see colour scale). The top two panels are equivalent to \cref{fig:results}(a,b) of the main text. The noise levels are divided by the appropriate responsivities (c.f. \cref{eq:Sn_exp} of the main text), as indicated in the y-axis labels.}\label{fig:AllPSDs}
\end{figure}
\begin{figure}[!ht]
	\includegraphics[width=.75\linewidth]{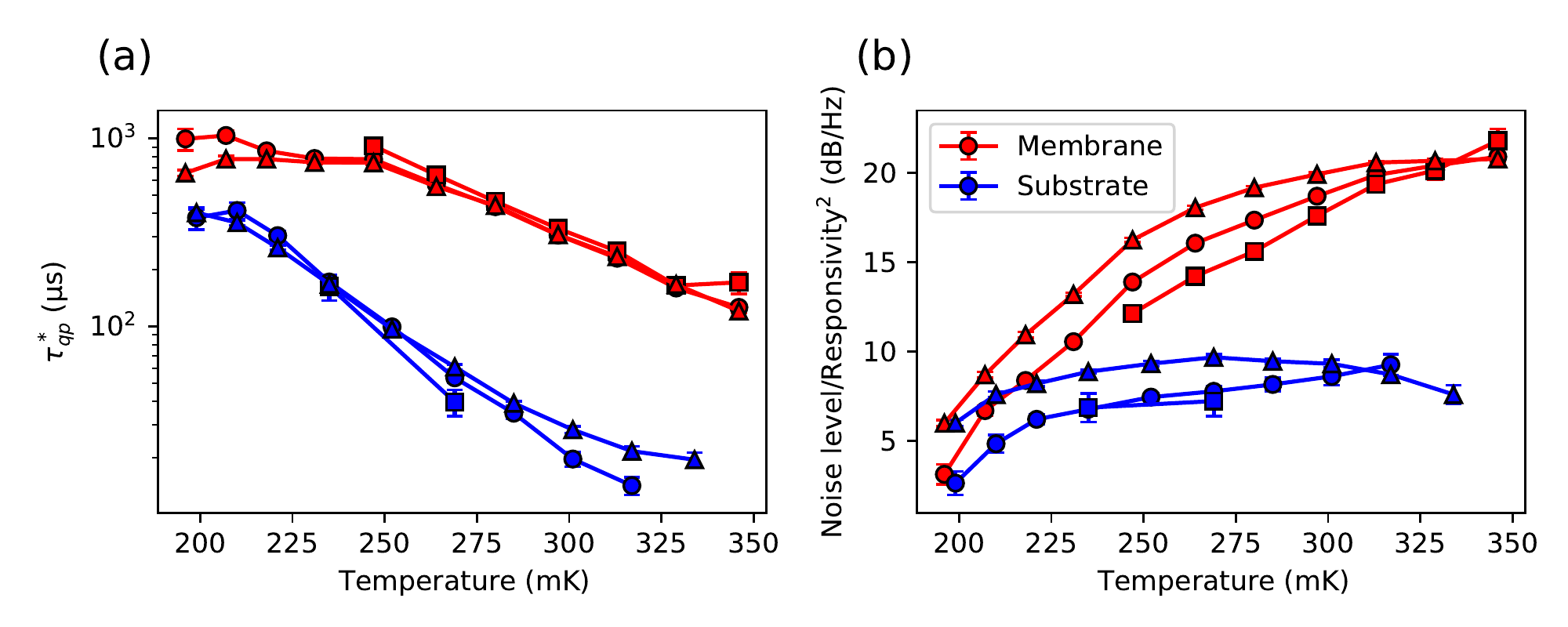}
	\caption{Lifetime (\textbf{a}) and noise level (\textbf{b}) from Lorentzian fits (\cref{eq:Sn} of the main text) to the spectra of \cref{fig:AllPSDs}. Red (blue) curves are for the membrane (substrate) resonator. The markers show which spectrum is fitted: ($\bigcirc$) is the cross-PSD ($S_{A,\theta}(\omega)$), ($\square$) is the amplitude PSD ($S_{A}(\omega)$) and ($\triangle$) is the phase PSD ($S_{\theta}(\omega)$). Error bars indicate the uncertainty from the fitting procedure and only fits with a relative uncertainty of less than \SI{16}{\percent} in lifetime are displayed. Before fitting to $S_{\theta}(\omega)$ and $S_A(\omega)$, the level at \SI{18}{\kilo\hertz} is subtracted from the spectrum, to eliminate amplifier noise.}\label{fig:AllFits}
\end{figure}
\newpage
\section{Pulse Rejection}

\begin{figure}[!ht]
	\includegraphics[width=.8\linewidth]{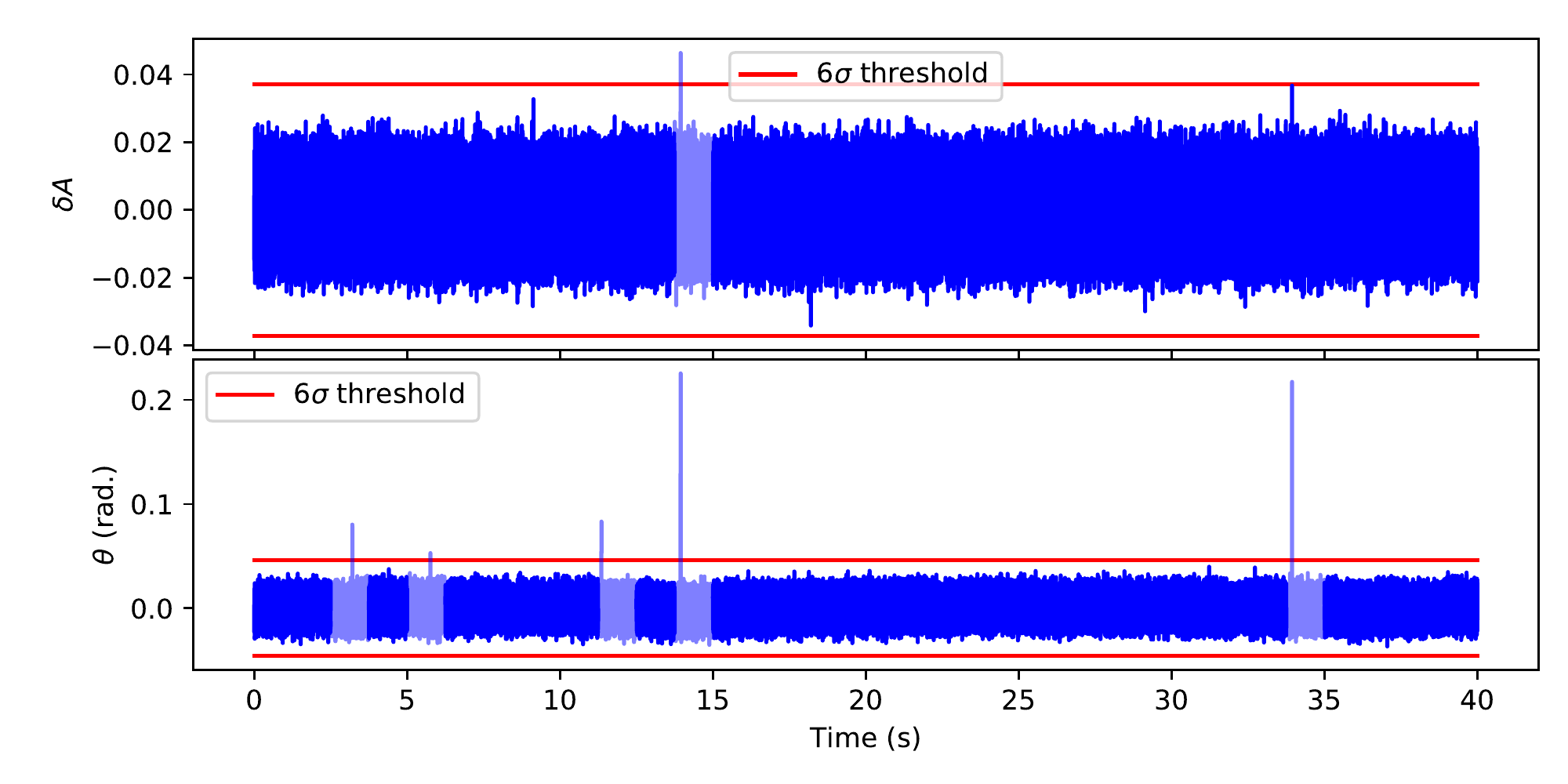}
	\caption{Example of pulse rejection of amplitude ($\delta A$) and phase ($\theta$), at \SI{141}{\milli\kelvin} and \SI{-99}{\decibel m} read-out power. The data is corrected for an offset. If a data segment contains a point outside of the $6\sigma$ interval, it is rejected, which is depicted by the semi-transparency of the data segment.}\label{fig:PulseRej}
\end{figure}

The time traces typically contain pulses, caused by external energy sources, such as cosmic rays \cite{Karatsu2019}. As the amplitude of such pulses is large compared to the noise, they have a dominating effect on the noise measurement and thus contaminate the Power Spectral Densities (PSDs). We filter the pulses in the following way.\\
First, the time trace is corrected for an offset by fitting a quadratic polynomial and subtracting it from the data. This is to correct for small temperature drifts during the measurement. Then, the time stream is cut into 32 equal segments. A segment is rejected if it contains a point outside a threshold. This threshold is set to $6\sigma_{min}$, where $\sigma_{min}$ is the minimal standard deviation of all the segments. This threshold ensures we filter pulses of \SI{6e2}{} and \SI{2e3}{} excess quasiparticles, for phase and amplitude, respectively. For the cross-PSD, the segment is rejected if the amplitude or the phase segment is rejected. An example of this procedure is shown in \cref{fig:PulseRej}.\\
For each non-rejected segment, the PSD is estimated via a periodogram with a Hamming window, after which they are averaged. Finally, the spectrum is down-sampled to 30 points per decade for visibility.\\

\section{Al Geometry Variations}
In the following sections, measurements of different devices will be discussed, which have not been presented in the main text. The measurement procedure and data analysis is the same. \Cref{tab:devices} shows an overview of the different parameters of the devices.
\begin{table}[!ht]
	\caption{An overview of the devices discussed with their distinguishing properties. The results presented in the main text are from device A1 and A2. The stars indicate that the values are measured on the same Al film, but on a strip of \SI{100}{\micro\meter} wide, instead of the Al strip width of the resonator.}\label{tab:devices}
	\begin{tabular}{ccccccccc}
		\hline\hline
		Device & Capacitor & Substrate    & Al thickness (nm) & Al width (µm) & Al length (mm) & $T_c$ (K)     & $\rho_N$ (\SI{}{\micro\ohm\centi\meter}) & RRR  \\ \hline
		A1     & IDC       & SiN membrane & 50             & 1.73       & 0.31        & 1.26          & 1.6                                   & 4.0  \\
		A2     & IDC       & SiN/Si substrate & 50             & 1.73       & 0.31        & 1.26          & 1.6                                   & 4.0  \\
		A3     & IDC       & SiN membrane & 25             & 1.73       & 0.31        & 1.35          & 2.4                                   & 2.7  \\
		B1     & IDC       & c-Sapphire   & 40             & 0.6        & 0.12 - 1.4  & 1.2*          & 0.7*                                  & 5.2* \\
		B2     & IDC       & c-Sapphire   & 40             & 1.5        & 0.12 - 1.4  & 1.2*          & 0.7*                                  & 5.2* \\
		C      & CPW       & c-Sapphire   & 150            & 0.92       & 0.53        & 1.12          & 0.4                                   & 9.3 \\ \hline\hline
	\end{tabular}
\end{table}

\subsubsection{Al Length}
\begin{figure}[!ht]
	\centering
	\includegraphics[width=\linewidth]{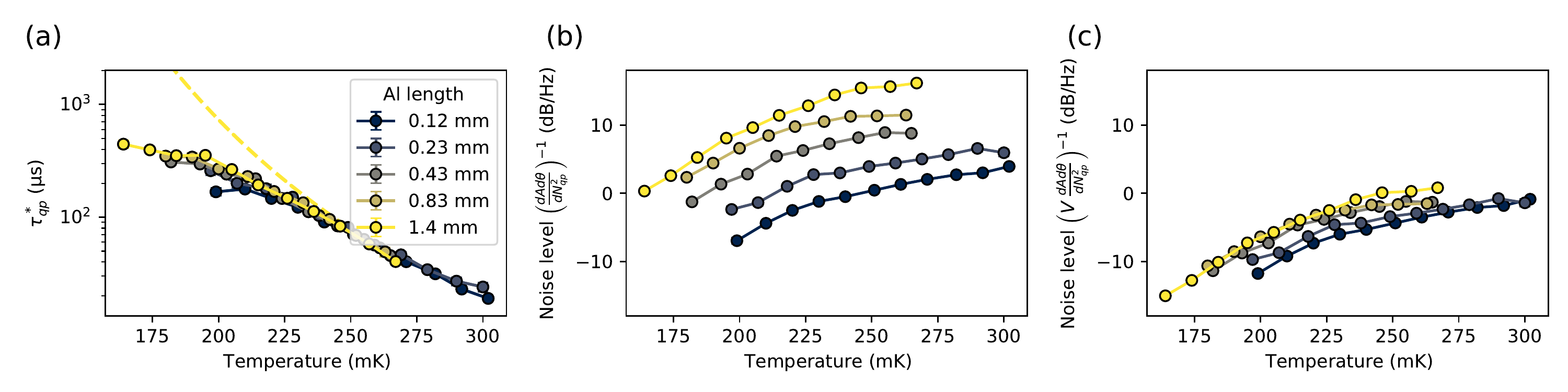}
	\caption{Lifetime (\textbf{a}) and noise levels (\textbf{b},\textbf{c}) for device B1, from $A$-$\theta$ cross-PSDs and responsivities (see \cref{eq:Sn_exp} of the main text) fitted with \cref{eq:Sn}, for different Al strip lengths given by the legend. Only fits with a relative error of \SI{>13}{\percent} in lifetime are shown and the error bars include statistical fitting uncertainties only. The dashed line is a fit to \cref{eq:tqpstar} for the \SI{1.4}{\milli\meter} data, with a fitted value $\tau_{esc}=\SI{0.14}{\nano\second}$. Panel \textbf{(c)} shows the same noise levels as \textbf{(b)}, but divided by the Al volume. Read-out powers vary from \SIrange{-110}{-113}{\decibel m}.}\label{fig:LengthVariation}
\end{figure}
We measured the quasiparticle fluctuation in device B1, which Al strip length variations from \SIrange{.12}{1.4}{\micro\meter}. \Cref{fig:LengthVariation} shows the result of the same procedure explained in the main text. \\
We do not see a clear dependence of the saturation lifetime on Al length. The increase in noise level with longer strip length, can partly be explained by noting that $S_{N_{qp}}(\omega) \propto V$. \Cref{fig:LengthVariation}(c) shows the noise levels when divided by Al volume.

\subsubsection{Al Width}

\begin{figure}[!ht]
	\centering
	\includegraphics[width=\linewidth]{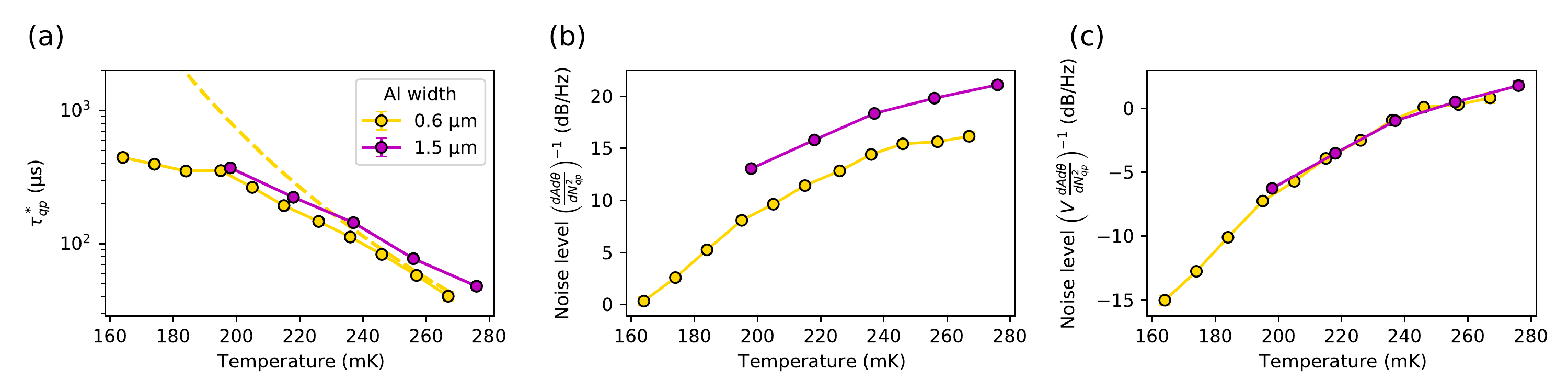}
	\caption{Lifetime (\textbf{a}) and noise levels (\textbf{b},\textbf{c}) from $A$-$\theta$ cross-PSDs and responsivities (see \cref{eq:Sn_exp}) fitted with \cref{eq:Sn}, for different Al strip widths given by the legend (devices B1 and B2). Only fits with a relative error of \SI{11}{\percent} in lifetime are shown and the error bars include statistical fitting errors only. The \SI{0.6}{\micro\meter} data and dashed line are the same as in \cref{fig:LengthVariation}. Panel \textbf{(c)} shows the noise levels divided by Al volume.}\label{fig:WidthVariation}
\end{figure}

In \cref{fig:WidthVariation}, we show a comparison in lifetime and noise levels, when the Al width is varied. Although the lifetimes for the \SI{1.5}{\micro\meter} strip are a bit higher (possibly due to a higher phonon escape time), no difference in saturation lifetime can be observed. The offset in noise levels can be explained by the larger Al volume, as \cref{fig:WidthVariation}(c) shows. 

\subsubsection{Al Thickness}
\begin{figure}[!ht]
	\centering
	\begin{subfigure}{0.7\linewidth}
		\includegraphics[width=\linewidth]{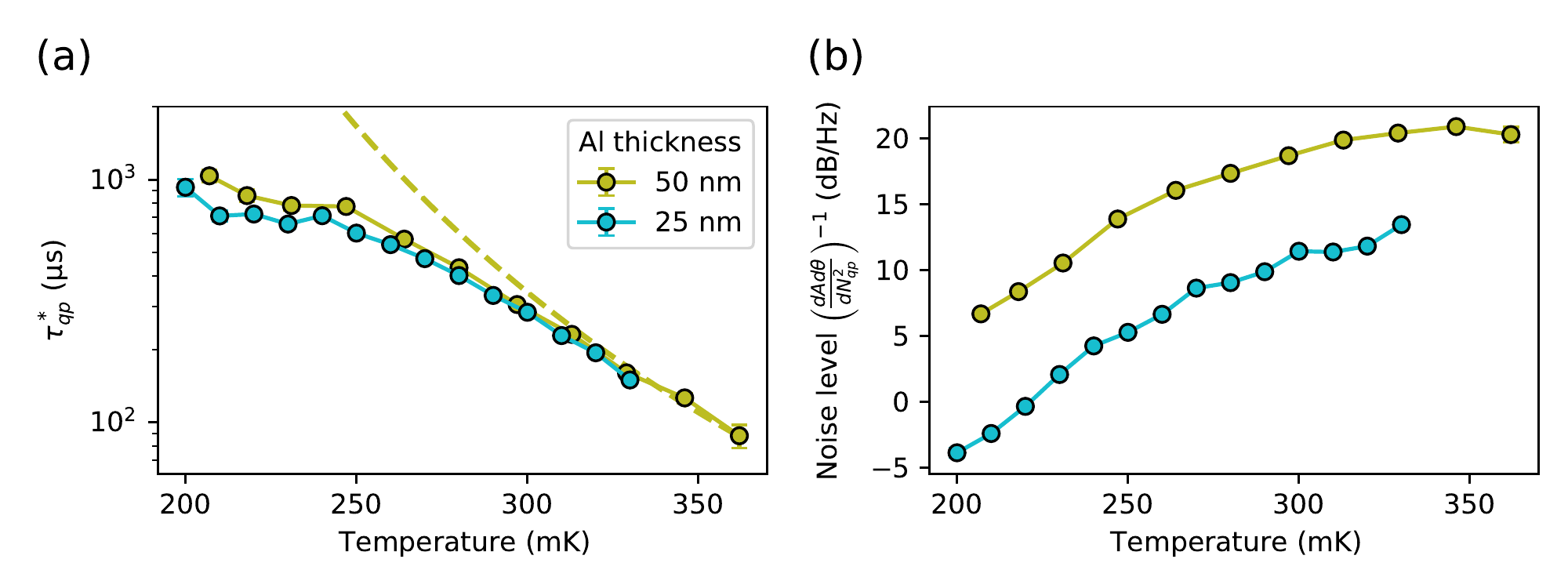}
	\end{subfigure}
	\begin{subfigure}{0.7\linewidth}
		\includegraphics[width=\linewidth]{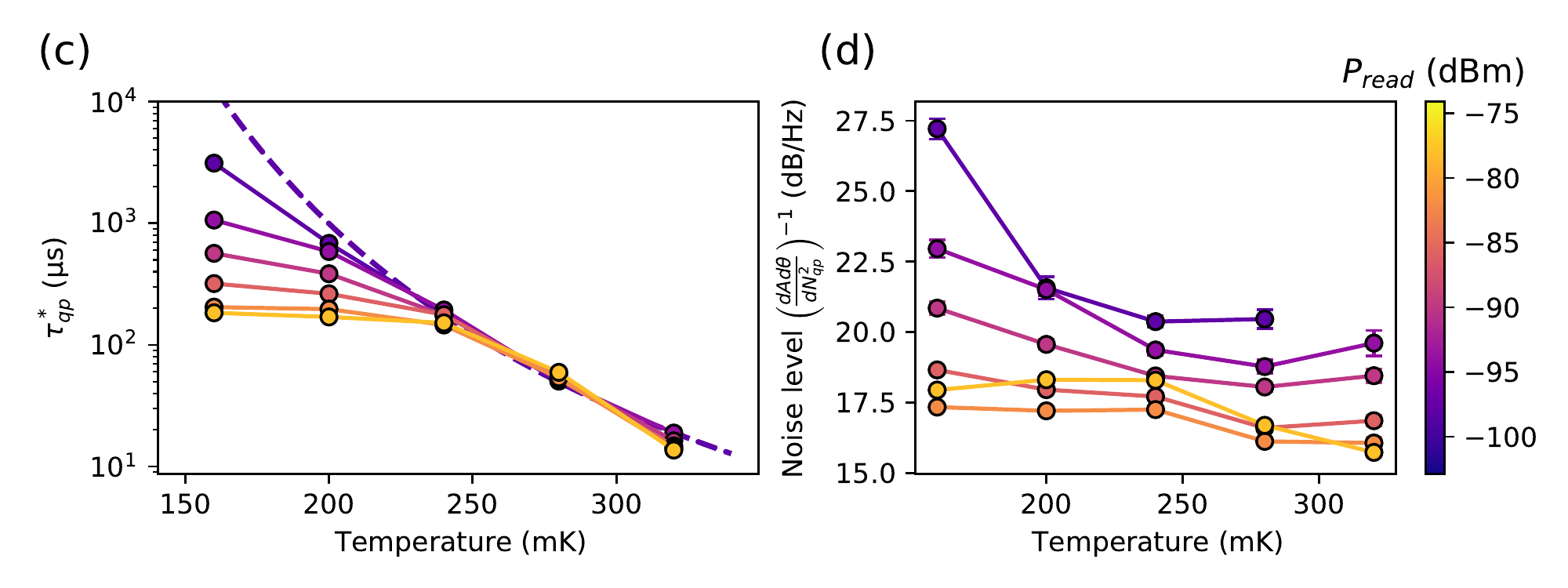}
	\end{subfigure}
	\caption{Lifetime (\textbf{a},\textbf{c}) and noise level (\textbf{b},\textbf{d}) from a fit of \cref{eq:Sn} in the main text, to the cross-PSD divided by responsivity squared (\cref{eq:Sn_exp} in the main text). \textbf{(a,b)}: Green curves are from \cref{fig:results}(c,d) in the main text (device A1) and the blue curves are for device A3 at \SI{-97}{\decibel m} read power. \textbf{(c,d)}: Lifetime and noise levels measured for device C, for a large variety of read powers, see colour scale. The dashed lines are fits of \cref{eq:tqpstar} of the main text, to the lowest read power data, only for temperatures \SI{>240}{\milli\kelvin}.}\label{fig:ThicknessVariations}
\end{figure}

The same device as discussed in the main text is also fabricated with an Al thickness of \SI{25}{\nano\meter} (device A3). A comparison of lifetime and noise level for the membrane resonators is given in \cref{fig:ThicknessVariations}(a,b). Only a slight difference in saturation lifetime can be observed. We would expect shorter lifetimes for device A3 than for A1, as the phonon escape time should be shorter. However, the \SI{25}{\nano\meter} film has a higher $T_c$ (as theoretically predicted in Ref. \cite{Chubov1969}), which also increases the lifetime (\cref{eq:tqp}). This diminishes the difference in lifetime we would expect from phonon trapping. To be precise, the phonon trapping factor is calculated to be 21 and 11 and the $T_c$ is measured to be \SI{1.255}{\kelvin} and \SI{1.35}{\kelvin} for the \SI{50}{\nano\meter} and \SI{25}{\nano\meter} resonators, respectively.\\
The \SI{8}{\decibel\per\hertz} difference in noise level can only partly be explained by the larger Al volume (\SI{3}{\decibel\per\hertz}). Differences in critical temperature (c.f. \SI{1.255}{\kelvin} and \SI{1.35}{\kelvin}) and resistivity (c.f. \SI{1.6}{\micro\ohm\centi\meter} 
and \SI{2.4}{\micro\ohm\centi\meter}, for the \SI{50}{\nano\meter} and \SI{25}{\nano\meter}, respectively) influences $\tau_0$ \cite{Kaplan1976}, while $\tau_0$ is taken constant here. This could be the cause for the large difference in noise level.\\

We also include the lifetimes and noise levels from a different resonator, which has a Al thicknesses of \SI{150}{\nano\meter}. In \cref{fig:ThicknessVariations}(c,d), we observe a high saturation lifetime for the lowest read power (\SI{\geq 3}{\milli\second}) and, interestingly, the reduction of noise level is not observed in this thicker resonator. This suggests that the cause of the noise level reduction is located either on the Al-substrate interface or the top surface, for example, by unpaired surface spins from native Al oxide, which is supported by the results of Ref. \cite{Barends2009a}. In contrast, the results of Ref. \cite{Fyhrie2020} show higher $\tau_{sat}$ for decreasing Al thickness.\\
The lifetime saturation that is still present, is most likely due to read power induced creation of excess quasiparticles as observed in Refs. \cite{deVisser2011,deVisser2012}, and supported by the fact that $\tau_{sat}$ decreases with increasing read power. Note that here, the read power is increased \SI{\sim 25}{\decibel} with respect to the other data presented up to this point.

\section{Microwave Readout Power Dependence}
\begin{figure}[!ht]
	\begin{subfigure}{\linewidth}
		\includegraphics[width=\linewidth]{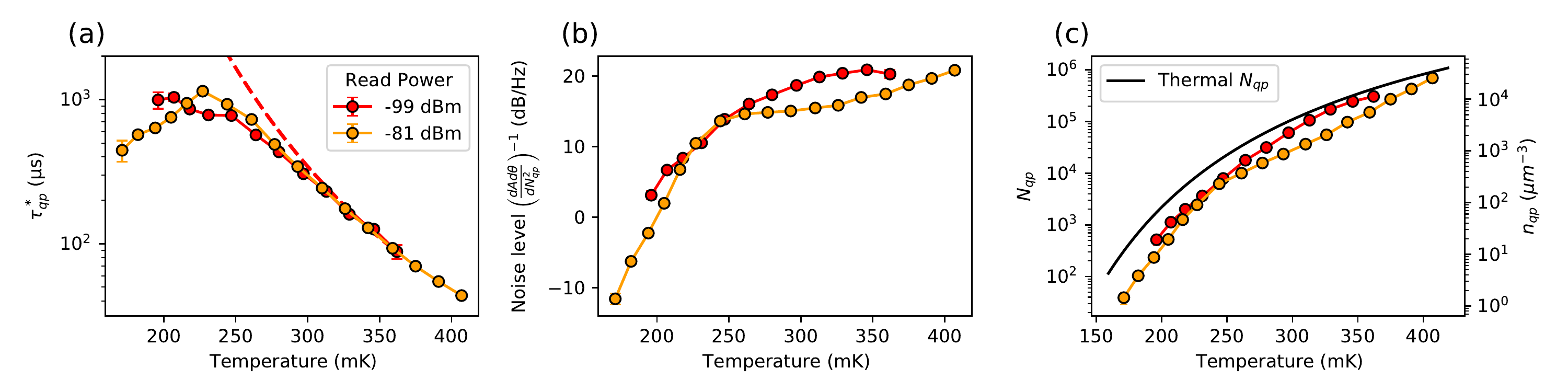}
	\end{subfigure}
	\begin{subfigure}{\linewidth}
		\includegraphics[width=\linewidth]{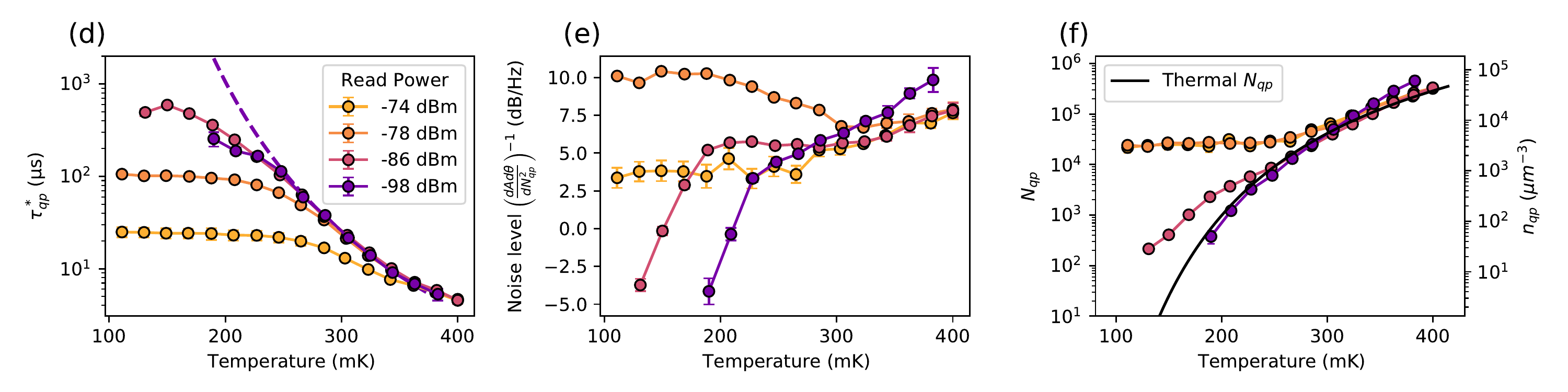}
	\end{subfigure}
	\caption{Lifetime (\textbf{a,d}), noise level (\textbf{b,e}) and calculated quasiparticle number (\textbf{c,f}) from a fit of \cref{eq:Sn} to the cross-PSD divided by responsivity squared (\cref{eq:Sn_exp}) at various read powers (see legend), for device A1 (\textbf{a-c}) and device B2 (\textbf{d-f}). The red curves in (a-c) are the same as in \cref{fig:results} and \cref{fig:Nqp} of the main text. The dashed lines are fits of \cref{eq:tqpstar} of the main text, to the lowest read power data with $\tau_{esc}$ as fitting parameter.}\label{fig:ReadPower}
\end{figure}

In the measurements discussed in the main text, we selected low read powers to limit read-out power effects, such as creation of excess quasiparticles \cite{deVisser2012,deVisser2014b}, redistribution of quasiparticles \cite{deVisser2014a} and non-linear kinetic inductance \cite{Swenson2013,Valenti2019}, which was recently reviewed in Ref. \cite{Thomas2020}. Here, we analyse the effect of increasing the read-out power.\\

\subsubsection{Peak in Lifetime}
\Cref{fig:ReadPower} shows how the results change when the read power is increased. In \cref{fig:ReadPower}(a,b) we observe a peak in lifetime around \SI{230}{\milli\kelvin} appearing at higher read power. This is also observed in Ref. \cite{Barends2009a}, where also trapping states are deemed to be the cause. We notice that $\tau_{sat}$ is now temperature dependent, implying that $N_t$ increases with decreasing temperature. From \cref{fig:ReadPower}(c), we see that this increase in read power does not result in excess (free) quasiparticles.\\
\subsubsection{Creation of Excess Quasiparticles}
\Cref{fig:ReadPower}(d-f) shows the noise behaviour at even larger read powers, measured in device B2. Upon increasing read power, the onset of the noise level reduction occurs to lower temperatures, after which it disappears. The saturation lifetime still decreases as read power increases, consistent with the creation of excess quasiparticles \cite{deVisser2012,deVisser2014b}, also observed in \cref{fig:ReadPower}(f). Indeed, we see that more excess quasiparticles are created when increasing the readout power from \SIrange{-98}{-78}{\decibel m}. Interestingly, we see that increasing the read power from \SIrange{-78}{-74}{\decibel m}, does not increase the amount of non-equilibrium quasiparticles, but does lower the saturation lifetime and noise level. This may be caused by a microwave redistribution effect on the responsivity \cite{deVisser2014a}, which we do not account for.\\
We conclude from these observations that a lifetime saturation can be caused by two phenomena. At high readout powers, excess quasiparticles are created, which saturates the free quasiparticle number ($N_{qp}$ in \cref{eq:Nqp} of the main text), and therefore also $\tau_{qp}$ (\cref{eq:tqp} of the main text). The quasiparticle lifetime is then limited by recombination events. This is consistent with Refs. \cite{deVisser2012,deVisser2011}. At low readout powers, the creation of excess quasiparticles is suppressed and the trapped quasiparticle number ($N_t$) becomes significant compared to $N_{qp}$. This limits the quasiparticle lifetime to \textit{on-trap} recombination events, while the free quasiparticle number ($N_{qp}$) stays thermal. These two phenomena can be distinguished by calculating $N_{qp}$ from the noise levels, as done in \cref{fig:ReadPower}(c,f) and \cref{fig:Nqp} from the main text.
\newpage
\section{Responsivity Measurement}
\begin{figure}[!ht]
	\begin{subfigure}{0.7\linewidth}
		\includegraphics[width=\linewidth]{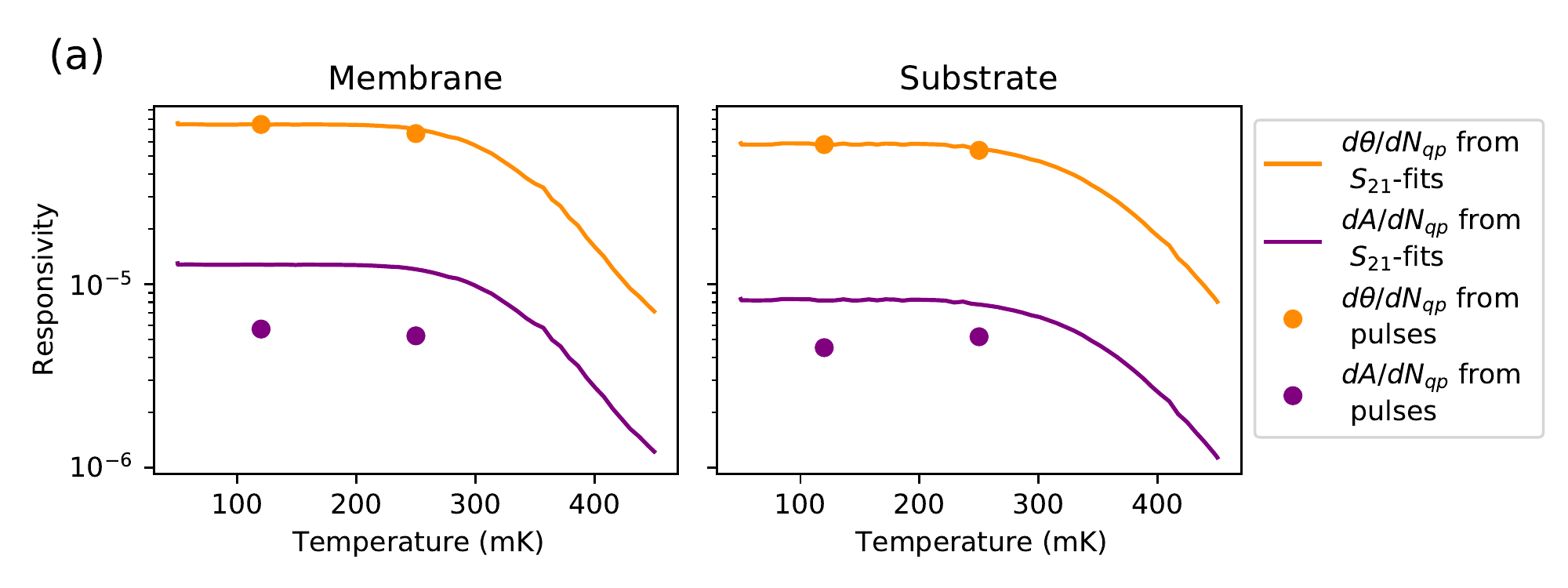}
	\end{subfigure}
	\begin{subfigure}{0.7\linewidth}
		\includegraphics[width=\linewidth]{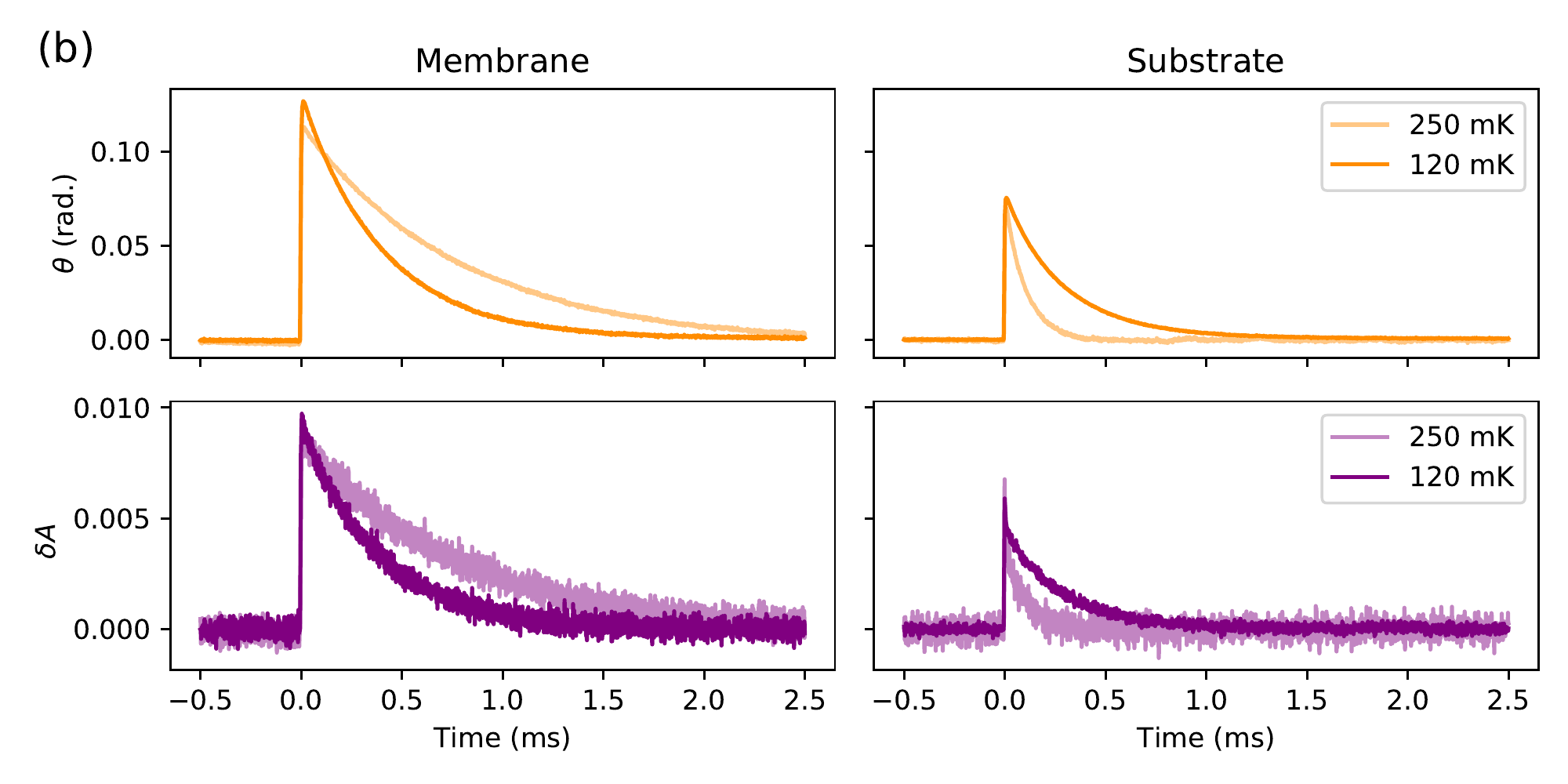}
	\end{subfigure}
	\caption{\textbf{(a)}: Amplitude and phase responsivities determined by two methods: from $S_{21}$-fits as explained in the text and from a single photon (\SI{1545}{\nano\meter}) response pulse height. \textbf{(b)}: Average amplitude and phase pulse response for a \SI{1545}{\nano\meter} photon.}\label{fig:PulseResp}
\end{figure}
We determine the amplitude and phase responsivities in a separate measurement of $S_{21}$ as a function of temperature. We fit Lorentzian curves to the $|S_{21}|^2$ dips to determine the resonance frequency ($f_0$) and the internal and coupling quality factor for each temperature and read power. We then linearly fit $1/Q_i(T)$ and $\delta f_0(T)/f_0(0)$ to $N_{qp}(T)$ at high temperatures ($>\SI{250}{\milli\kelvin}$), where quasiparticle loss dominates. Here, $\delta f_0 = f_0(T) - f_0(0)$, where $f_0(0)$ is the resonance frequency at \SI{50}{\milli\kelvin} and $N_{qp}(T)$ is calculated from \cref{eq:Nqp} from the main text. The amplitude and phase responsivities are calculated as $dA/dN_{qp} = -2Q(T)\frac{d(1/Q_i)}{dN_{qp}}$ and $d\theta/dN_{qp} = 4Q(T)\frac{d(\delta f_0/f_0(0))}{dN_{qp}}$, respectively \cite{deVisser2014}. This method does not take read power effects into account, but it is executed at the same read power as the noise measurement.\\

\Cref{fig:PulseResp}(a) shows the calculated responsivities along with a different method to calculate the responsivity. This second method is based on the single photon response of the resonator, which is used as an MKID \cite{Day2003}. When a photon is absorbed in the Al, it will break thousands of Cooper-pairs, resulting in an excess quasiparticle number, $\delta N_{qp}$, and therefore a measurable signal in $A$ and $\theta$. The responsivity is calculated as,
$$
\frac{dX}{dN_{qp}} = X_{max}\left(\delta N_{qp}\right)^{-1}=X_{max}\left(\eta_{pb}\frac{\hbar\omega_\gamma}{\Delta}\right)^{-1},
$$
where $X$ is either $\theta$ or $A$ and $X_{max}$ is the maximum measured value. $\eta_{pb}$ is the pair-breaking efficiency, which we here set to \SI{.40}{} and \SI{.31}{} for the membrane and substrate resonator respectively \cite{deVisser2021a}. These values are chosen such that the two calculation methods coincide for the phase variable at \SI{120}{\milli\kelvin} and should be compared to the theoretical maximum value of \SI{0.59}{} \cite{Guruswamy2014}. $\hbar\omega_\gamma$ is the \SI{1545}{\nano\meter}-photon energy, i.e. \SI{0.8}{\electronvolt}. For $\Delta$, we use the BSC-relation $2\Delta=3.52k_BT_c$.\\
Experimentally, a \SI{1545}{\nano\meter} light source illuminates the middle part of the chip. The laser is attenuated to the point where single photon absorptions can be detected, without much overlap. A time stream of \SI{40}{\second} is measured at \SI{1}{\mega\hertz}, with a read frequency equal to the resonance frequency of the resonator and power $P_{read}=\SI{-85}{\decibel m}$ and \SI{-83}{\decibel m} for on and off membrane, respectively. The $I$ and $Q$ signals from the $IQ$-mixer are translated to $A$ and $\theta$ with the use of the resonance circle, which is measured earlier.\\
The average of multiple (typically a thousand) single photon pulses is presented in \cref{fig:PulseResp}(b). A detailed description of the experiment and results of phonon trapping effects to the pulse amplitudes will be published separately \cite{deVisser2021a}.\\

In \cref{fig:PulseResp}(a) both methods show that the responsivity does not decrease over the temperature range where the noise level reduction is observed (c.f. \cref{fig:results}(c,d)). We conclude that the noise level reduction is caused by quasiparticle dynamics, and not by a reduction in responsivity (i.e. by $S_{N_{qp}}$ and not by the last factor in  \cref{eq:Sn_exp} of the main text).\\

\end{document}